\documentclass[a4paper,fleqn,usenatbib]{mnras}
\usepackage{newtxtext,newtxmath}
\usepackage[T1]{fontenc}
\usepackage{ae,aecompl,bm,booktabs}
\usepackage{amsmath,color,xspace,multirow,longtable,graphicx,bigints,comment}
\usepackage{tabularx}
\usepackage{pdflscape}

\newcommand\textlcsc[1]{\textsc{\MakeLowercase{#1}}}
\newcommand{\cii}{[C\,II]\xspace}
\newcommand{\oiii}{[O\,III]\xspace}

\title[Evidence for extended gaseous reservoirs at $z\sim2$]{Evidence for extended gaseous reservoirs around AGN at cosmic noon from ALMA CO(3-2) observations}

\author[G. C. Jones et al.]{
G. C. Jones$^{1}$\thanks{E-mail: gareth.jones@physics.ox.ac.uk},
R. Maiolino$^{2,3,4}$, 
C. Circosta$^{4,5}$,
J. Scholtz$^{2,3}$, 
S. Carniani$^{6}$,
Y. Fudamoto$^{7,8}$
\\
$^{1}$Department of Physics, University of Oxford, Denys Wilkinson Building, Keble Road, Oxford OX1 3RH, UK\\
$^{2}$Cavendish Laboratory, University of Cambridge, 19 J. J. Thomson Ave., Cambridge CB3 0HE, UK\\
$^{3}$Kavli Institute for Cosmology, University of Cambridge, Madingley Road, Cambridge CB3 0HA, UK\\
$^{4}$Department of Physics \& Astronomy, University College London, Gower Street, London WC1E 6BT, UK\\
$^{5}$European Space Agency (ESA), European Space Astronomy Centre (ESAC), Camino Bajo del Castillo s/n, 28692 Villanueva de la Ca\~{n}ada, Madrid, Spain\\
$^{6}$Scuola Normale Superiore, Piazza dei Cavalieri 7, I-56126 Pisa, Italy\\
$^{7}$Waseda Research Institute for Science and Engineering, Faculty of Science and Engineering, Waseda University, 3-4-1 Okubo, Shinjuku, Tokyo 169-8555, Japan\\
$^{8}$National Astronomical Observatory of Japan, 2-21-1, Osawa, Mitaka, Tokyo, Japan
}
\begin{document}
\label{firstpage}
\pagerange{\pageref{firstpage}--\pageref{lastpage}}
\maketitle

\begin{abstract}
Gaseous outflows are key phenomena in the evolution of galaxies, as they affect star formation (either positively or negatively), eject gas from the core or disk, and directly cause mixing of pristine and processed material. Active outflows may be detected through searches for broad spectral line emission or high-velocity gas, but it is also possible to determine the presence of past outflows by searching for extended reservoirs of chemically enriched molecular gas in the circumgalactic medium (CGM) around galaxies. In this work, we examine the CO(3-2) emission of a set of seven $z\sim2.0-2.5$ AGN host galaxies, as observed with ALMA. Through a three-dimensional stacking analysis we find evidence for extended CO emission of radius $r\sim13$\,kpc. We extend this analysis to the HST/ACS i-band images of the sample galaxies, finding a complex small-scale ($r<10$\,kpc) morphology but no robust evidence for extended emission. In addition, the dust emission (traced by rest-frame FIR emission) shows no evidence for significant spatial extension. This indicates that the diffuse CO emission revealed by ALMA is morphologically distinct from the stellar component, and thus traces an extended reservoir of enriched gas. The presence of a diffuse, enriched molecular reservoir around this sample of AGN host galaxies at cosmic noon hints at a history of AGN-driven outflows that likely had strong effects on the star formation history of these objects.
\end{abstract}

\section{Introduction}\label{intro}

The diffuse circumgalactic medium (CGM) surrounding a galaxy is a record of its recent past and an indicator of its future. This spatially extended ($\mathcal{O}[10-100$\,kpc]; e.g., \citealt{proc17,tuml17,fuji19}) material, which acts as a major reservoir of fuel for star formation, may include streaming gas from large-scale gaseous filaments (e.g., \citealt{arri18,umeh19}), as well as processed materials ejected from the galaxy itself. Gas may be expelled from the central source by starburst-driven winds (e.g., \citealt{gall18,spil18,jone19}) or by feedback from an active galactic nucleus (AGN; e.g., \citealt{Fluetsch2019,bisc19,Lutz20,trav20,vayn21}), hinting at energetic processes in the center of the galaxy or its disk. This expelled gas may form an extended gaseous reservoir, which can be explored using multiple tracers (e.g., HI\,21\,cm, Ly$\alpha$, optical nebular lines, \cii\,158\,$\mu$m, CO rotational transitions).

The most direct has been the HI line, which traces the atomic medium. Using MeerKAT and its predecessor, studies have revealed a diffuse component of HI gas in local starburst galaxies that extends beyond the stellar disk (e.g., \citealt{luce15,ianj22}). The kinematics of these extended gaseous reservoir differ from those of the disk, suggesting that the gas was expelled by past starburst-driven winds.

In addition to the several detections via absorption features towards the line of sight of quasars in the background of the CGM of galaxies \citep{Werk16,tuml17}, some of the most robust detections of the CGM in emission have been through VLT/MUSE observations of Ly$\alpha$ emission in $z>2$ AGN (e.g., \citealt{gino18,drak20,sand21}), where halos up to $\sim170$\,pkpc have been discovered. However, the resonant nature of Ly$\alpha$ makes the process of extracting the morphology and kinematics of the underlying CGM non-trivial. In addition, a significant fraction of the CGM mass may be in the cold phase. Therefore, in order to trace the true gas distribution, including the cold component, it is necessary to use other direct tracers, such as \cii (e.g., \citealt{zane18}) or CO (e.g., \citealt{bola13}).

Due to its low excitation energy, \cii emerges from a number of gas phases (i.e., warm ionized, warm/cold diffuse atomic, and warm dense molecular; \citealt{pine13}). Local observations have revealed a weak, diffuse component of \cii emission around some galaxies (e.g., \citealt{madd93}) which likely trace the gas in the atomic HI disk rather than the molecular distribution. At high-redshift, extended gas reservoirs have been detected in both individual sources and stacks of \cii emission in star-forming main sequence (MS; e.g., \citealt{noes07}) galaxies at $z>4$ (e.g., \citealt{fuji19,fuji20}), suggesting significant enrichment of the CGM after only $<1.5$\,Gyr. This \cii emission is clearly larger than the ALMA synthesized beam, and may extend further than the underlying rest-frame UV emission (tracing young stars). The extended gas reservoirs are taken as evidence for starburst-driven outflows (e.g., \citealt{pizz20,gino20,herr21}), which could eject processed material into the CGM (e.g., \citealt{heck90}). The galaxies in these \cii studies were chosen to exclude galaxies with type 1 AGN, and the extended \cii emission was found to be more evident in sources with higher star formation rates (SFRs), strengthening the star formation-driven outflow hypothesis. 

While evidence for gaseous outflows and extended gas reservoirs in SFGs at $z>4$ have been well-studied with ALMA, it is important to also search for these signatures in AGN host galaxies at cosmic noon (i.e., $z\sim2$; \citealt{Ginolfi17,li21,debr22}). One recent survey in this epoch is SUPER (a SINFONI Survey for Unveiling the Physics and Effect of Radiative feedback; \citealt{circ18}), which used VLT/SINFONI to target 39 x-ray--selected AGN host galaxies at $z\sim2.0-2.5$. SINFONI observations of the \oiii$\lambda$5007 line in 21 of these sources reveal evidence for outflows in every observed source \citep{kakk20}, with large outflow velocities ($\sim10^3$\,km\,s$^{-1}$). These sources were chosen to lie in well-studied survey fields (i.e., COSMOS; \citealt{scov07a,scov07b}, XMM-XXL north field; \citealt{pier16}), allowing for the derivation of stellar masses (log$_{10}$(M$_*$/M$_{\odot}$)= [9.59-11.21]), star formation rates (SFR=[25-680]\,M$_{\odot}$\,year$^{-1}$), and AGN bolometric luminosities (log(L$_{\mathrm{bol}}$/erg\,s$^{-1}$)=[44.3-47.9]) using CIGALE (\citealt{boqu19}). These sources were then observed in CO(3-2) emission with ALMA \citep{circ21}, resulting in 11 detections (from 27 targets). 

In this work, we perform a stacking analysis of the ALMA CO(3-2) data in order to investigate whether these energetic, outflowing sources exhibit extended reservoirs of molecular gas. We begin by describing the properties of the observations and our data reduction and image creation process in Section \ref{obsdat}. The stacking analysis, radial profile extraction technique, and radial fitting details are presented in Section \ref{coanaly}. These ALMA CO(3-2) findings are discussed in Section \ref{disc}. We conclude in Section \ref{conc}.

We assume a standard concordance cosmology ($\Omega_{\Lambda}$,$\Omega_m$,h)=(0.7,0.3,0.7) throughout. Between $z=2.0-2.5$, $1''$ corresponds to 8.370-8.071\,kpc, respectively.

\section{Observations and Data Reduction}\label{obsdat}
In this project, we wish to examine $z\sim2.1-2.5$ QSO host galaxies observed in CO(3-2) emission as part of ALMA projects 2016.1.00798.S and 2017.1.00893.S (PI: V. Mainieri; \citealt{circ21}). 

Of the 28 sources observed as part of the two ALMA programs, we exclude all sources not detected in CO(3-2) emission (17 sources) and those with evidence of close companions based on ALMA CO(3-2) and rest-frame FIR images (3 sources: CID\_971, CID\_1215, and CID\_1253). Following \citet{circ21}, we exclude CID$\_$1057 due to an uncertain redshift. While a weak detection of CO(3-2) from X\_N\_6\_27 is reported in \citet{circ21}, deeper observations show that the emission is intrinsically fainter, and at a different redshift (see Appendix \ref{627sec}). Because of this, we exclude X\_N\_6\_27, leaving a final sample of seven galaxies. The reported properties of these galaxies are presented in Table \ref{basictab}. 

This subsample features a small redshift range ($2.2\lesssim z\lesssim2.5$) and only contains broad line (BL) AGN. When available, the SFRs and stellar masses of each object place them around the star-forming main sequence (SFR$\sim 48-686$\,M$_{\odot}$\,year$^{-1}$, log$_{10}$(M$_*$/M$_{\odot}$)$\sim10.30-11.21$; \citealt{circ21}).

Each of these ALMA projects was executed in Cycle 4 or 5, so their data are now public. Since our science goals require homogeneity of the data, we chose to reanalyze each dataset, beginning at the visibilities. The raw science data models (SDMs) were passed through a calibration pipeline created by ALMA staff, resulting in calibrated measurement sets (MSs) for each individual observation. These included multiple sources: flux/bandpass calibrators, phase calibrators, the target source, and occasionally a pointing calibrator. Because of this, we created new MSs containing only calibrated visibilities of the target source (CASA \textlcsc{split}). Some sources were observed over two executions, so their individual MSs were joined into one (CASA \textlcsc{concat}). With the calibrated MSs of each source in hand, we created an imaging pipeline, in order to ensure homogeneity of the resulting images. 

The first step of the imaging pipeline is to determine which channels may contain CO(3-2) emission. While we may adopt the CO(3-2) FWHM as derived from previous ALMA data (\citealt{circ21}), it is possible that the spatially extended emission features a different spectral width. That is, since these AGN host galaxies feature small-scale outflows \citep{kakk20}, they may feature a spectral component that is low-amplitude but broad in velocity space, as seen in other AGNs with outflows (e.g., \citealt{bisc19}).

To take this idea into account, we first assumed the CO(3-2) redshift reported by \citet{circ21} ($z_{CO}$ in Table \ref{basictab}) and took a range of $\pm800$\,km\,s$^{-1}$. This value is based on the largest CO(3-2) full-width at half-maximum (FWHM) reported in \citet{circ21}: $810\pm93$\,km\,s$^{-1}$ for CID\_1253 (a triple-component system excluded from this work). This conservative estimate ensures that we exclude all CO(3-2) emission from our `line-free' channels.

We then use the CASA task \textlcsc{uvcontsub} to fit the source visibilities with a first order model using all line-free channels and to subtract this model from the data, resulting in a `continuum-free' MS. These continuum-free visibilites are imaged using CASA tclean in `cube' mode with natural weighting and $0.2''$ cells to create a `dirty' line cube. The channel width is kept to its intrinsic value of $\sim7.812$\,MHz (roughly 25\,km\,s$^{-1}$). Using the CASA task \textlcsc{imsubimage}, we then trim the channel range to only include the sideband containing CO emission and to exclude the uncalibrated channels at the edge of each SPW. The RMS noise level of the channel of this cube is calculated using sigma clipping (using astropy \textlcsc{sigma\_clipped\_stats} with $S/N_{upper}=3$; \citealt{astr13,astr18}). The tclean task is then repeated, but the image is cleaned down to $3\times$ the RMS of the `dirty' image, creating a final `clean' line cube. The rest frequency of the line is set to the value found by \citet{circ21}. 

This process resulted in seven continuum-free data cubes with similar synthesized beam sizes ($\sim1''$) and RMS noise levels ($\sim0.3-0.6$\,mJy\,beam$^{-1}$; see Table \ref{basictab}).

\section{CO Cube Analysis}\label{coanaly}

With seven homogeneous CO(3-2) data cubes in hand, we proceed by stacking them in image-space and creating moment zero maps of each stacked cube (Section \ref{stamed}). The brightness distribution of each map is then analyzed by extracting radial profiles (Section \ref{RPE}). We use a simple 2-D Gaussian model to determine whether there is greater evidence for a single resolved source or a compact source surrounded by an extended component (Section \ref{HALOMOD}).

\begin{table*}
\centering
\caption{Properties of SUPER galaxies, as reported by \citet{circ18,circ21}. We also note whether each source has available HST/SC i-band photometry and whether the galaxy features a narrow line (NL) or broad line (BL) AGN (see Section \ref{hstsect}). For the seven objects included in our stacking analysis (Section \ref{stamed}), we also note the mean synthesized beam and RMS noise level per channel at the expected redshift of CO emission. $a$: We exclude all sources with evidence of close, significant companions (3 sources: CID\_971, CID\_1215, and CID\_1253). $b$: See Appendix \ref{627sec} for details of why this source is excluded.}
\label{basictab}
\begin{tabular}{l|ccc|ccc|cc|cc}
Source & RA & Dec & $z_{spec}$ & CO & $z_{CO}$ & ALMA & HST & AGN & Mean Synthesized & RMS$_{\mathrm{chan}}$\\ 
 & & & & & & Project & & Type & Beam & [mJy\,beam$^{-1}$]\\
\hline
CID$\_$1605 & 09:59:19.82 & +02:42:38.73 & 2.121 & n & $-$ & 2017.1.00893.S & y & BL & $-$ & $-$\\
CID$\_$357 & 09:59:58.02 & +02:07:55.10 & 2.136 & n & $-$ & 2017.1.00893.S & y & BL & $-$ & $-$\\
LID$\_$3456 & 09:58:38.40 & +01:58:26.83 & 2.146 & n & $-$ & 2017.1.00893.S & y & BL & $-$ & $-$\\
CID$\_$1253 &  10:01:30.57 & +02:18:42.57 & 2.147 & y$^a$ & $2.1508(4)$ & 2016.1.00798.S & y & NL & $-$ & $-$\\
X$\_$N$\_$102$\_$35 & 02:29:05.94 & -04:02:42.99 & 2.190 & n & $-$ & 2016.1.00798.S & n & BL & $-$ & $-$\\
CID$\_$38 & 10:01:02.83 & +02:03:16.63 & 2.192 & n & $-$ & 2017.1.00893.S & y & NL & $-$ & $-$\\
CID$\_$346 & 09:59:43.41 & +02:07:07.44  & 2.219 & y & $2.2198(1)$ & 2016.1.00798.S & y & BL & $(1.22''\times1.12''),51^{\circ}$ & 0.6\\
CID$\_$337 & 09:59:30.39 & +02:06:56.08 & 2.226 & n & $-$ & 2016.1.00798.S & y & NL & $-$ & $-$\\
CID$\_$852 &  10:00:44.21 & +02:02:06.76 & 2.232 & n & $-$ & 2017.1.00893.S & y & NL & $-$ & $-$\\
X$\_$N$\_$104$\_$25 & 02:30:24.46 & -04:09:13.39 & 2.241 & n & $-$ & 2016.1.00798.S & n & BL & $-$ & $-$\\
X$\_$N$\_$44$\_$64 & 02:27:01.46 & -04:05:06.73 & 2.252 & y & $2.245(1)$ & 2016.1.00798.S & n & BL & $(1.23''\times1.14''),-63^{\circ}$ & 0.7\\
CID$\_$1205 & 10:00:02.57 & +02:19:58.68 & 2.255 & n & $-$ & 2017.1.00893.S & y & BL & $-$ & $-$\\
X$\_$N$\_$6$\_$27 & 02:23:06.32 & -03:39:11.07 & 2.263 & n$^b$ & $-$ & 2016.1.00798.S & n & BL & $-$ & $-$\\
CID$\_$467 &  10:00:24.48 & +02:06:19.76 & 2.288 & n & $-$ & 2016.1.00798.S & y & BL & $-$ & $-$\\
X$\_$N$\_$81$\_$44 & 02:17:30.95 & -04:18:23.66 & 2.311 & y & $2.2950(3)$ & 2016.1.00798.S & n & BL & $(1.38''\times1.14''),66^{\circ}$ & 0.5\\
X$\_$N$\_$128$\_$48 & 02:06:13.54 & -04:05:43.20 & 2.323 & n & $-$ & 2016.1.00798.S & n & BL & $-$ & $-$\\
LID$\_$206 & 10:01:15.56 & +02:37:43.44 & 2.330 & y & $2.3326(2)$ & 2016.1.00798.S & y & BL & $(1.53''\times1.17''),-70^{\circ}$ & 0.4\\
LID$\_$1289 & 09:59:14.65 & +01:36:34.99 & 2.408 & n & $-$ & 2016.1.00798.S & y & NL & $-$ & $-$\\
CID$\_$247 & 10:00:11.23 & +01:52:00.27 & 2.412 & n & $-$ & 2017.1.00893.S & y & BL & $-$ & $-$\\
X$\_$N$\_$53$\_$3 & 02:20:29.84 & -02:56:23.41 & 2.434 & y & $2.433(3)$ & 2016.1.00798.S & n & BL & $(1.38''\times1.19''),62^{\circ}$ & 0.5\\
CID$\_$2682 &  10:00:08.81 & +02:06:37.66 & 2.435 & n & $-$ & 2016.1.00798.S & y & NL & $-$ & $-$\\
LID$\_$1852 & 09:58:26.57 & +02:42:30.22 & 2.444 & n & $-$ & 2017.1.00893.S & y & NL & $-$ & $-$\\
CID$\_$166 & 09:58:58.68 & +02:01:39.22 & 2.448 & y & $2.461(1)$ & 2016.1.00798.S & y & BL & $(1.30''\times0.88''),65^{\circ}$ & 0.3\\
CID$\_$451 &  10:00:00.61 & +02:15:31.06 & 2.450 & y & $2.4450(3)$ & 2017.1.00893.S & y & NL & $(1.61''\times1.39''),-55^{\circ}$ &0.3\\
CID$\_$1215 & 10:00:15.49 & +02:19:44.58 & 2.450 & y$^a$ & $2.446(1)$ & 2017.1.00893.S & y & BL & $-$ & $-$\\
CID$\_$971 & 10:00:59.45 & +02:19:57.44 & 2.473 & y$^a$ & $2.4696(1)$ & 2016.1.00798.S & y & NL & $-$ & $-$\\
CID$\_$970 & 10:00:56.52 & +02:21:42.35 & 2.501 & n & $-$ & 2017.1.00893.S & y & NL & $-$ & $-$\\
\end{tabular}
\end{table*}		

\subsection{Stacking Method}\label{stamed}
While each of the sources in our subsample have been detected in CO(3-2) emission, most of them have relatively low significance (i.e., $<5\sigma$ in integrated intensity). Since we wish to search for low-level emission, we require higher S/N, and thus must perform stacking of each data cube. 

To begin, we initialize an empty cube with spatial dimensions of 100$\times$100 pixels ($20''\times20''$, with $0.2''$/pixel) and a spectral axis that covers from -3000$<$v$<$3000\,km\,s$^{-1}$ in 240 channels of width 25\,km\,s$^{-1}$. These spatial and spectral pixel scales are chosen to replicate those of the input cubes. Each CO(3-2) cube is then read, and a 3-D cutout is made of the central 100$\times$100 spatial pixels. 

We consider four possible weighting schemes when stacking our sample of data cubes:

\vspace{1mm}
\textit{Inverse Variance} (InvV): The mean RMS of each cube is found, and the weight of cube \textit{n} is set to this inverse square of this value (i.e., the inverse variance): $w_n=1/\sigma_n^2$. In this way, noisier cubes are down-weighted.

\textit{Normalization} (Norm): The integrated CO flux density of each source ($I_{CO}\equiv S\Delta v$) is taken from \citet{circ21}, and each weight is set to the inverse of this value: $w_n=1/I_{CO,n}$. This ensures that strong detections do not dominate the stack.  

\textit{Inverse Variance \& Normalization} (InNo): The two previous weighting schemes are combined: $w_n=1/(\sigma_n^2\times I_{CO,n})$.

\textit{No weighting} (None): No weighting is performed ($w_n=1$). A stack using this weighting scheme is the same as a basic average of all cubes, with no assigned weights.
\vspace{1mm}

While our imaging pipeline and sample selection is designed to maximize the spatial match between the stacked cube and each input cube, the spectral scale is necessarily offset between each input cube. That is, while the channel width is uniform in frequency for all cubes, each is tuned to a different central frequency, and thus has a different velocity width. Because the velocity bins of each spectrum are different, we distribute the flux of each input spectrum into the bins of the stacked cube by calculating the amount of overlap. Considering a bin in the stacked cube with a central velocity of $v_a$ and width $\Delta v_a$, and an input bin with central velocity $v_b$ and width $\Delta v_b$, we calculate the fraction of the input bin that is inside the bin of the stacked cubes as:
\begin{equation}
F(v_a,v_b)=\frac{min\left(v_a+\frac{\Delta v_a}{2},v_b+\frac{\Delta v_b}{2}\right)-max\left(v_a-\frac{\Delta v_a}{2},v_b-\frac{\Delta v_b}{2}\right)}{\Delta v_b}
\end{equation}
where $F(\nu_a,\nu_b)$ is limited to [0,1].

For a set of N input data cubes, each with flux density $S_j(x_i,y_j,v_B)$ and weight $w_n$, we determine the stacked spectrum $S_{stack}(x_i,y_j,v_A)$ by using:
\begin{equation}\label{3dstack}
S_{stack}(x_i,y_j,v_A)=\frac{\Sigma_{v_B}\Sigma_{n=1}^{N}S_n(x_i,y_j,v_B)F(v_A,v_B)w_n}{\Sigma_{n=1}^{N}w_n}
\end{equation}
Note that this form is similar to past stacking analyses (e.g., \citealt{delh13,bisc19,gino20,joll20}). The redshift of the stacked cube is set to be the average of the input galaxy redshifts (i.e., $z_{stack}=2.337$).

In each case we assume that the galaxy is centred at the phase centre, which was selected based on rest-frame UV data. Due to dust obscuration, it is quite likely that the CO and UV distributions differ, and the true spatial centroid may be displaced by a small distance (e.g., \citealt{chen17,kaas20}). As an alternative, it is theoretically possible to fit each ALMA CO(3-2) moment zero map with a 2-D Gaussian model, and realign each cube to the resulting spatial centroid. But since multiple sources in this sample feature low integrated CO S/N values ($\sim3-4$), this approach is likely to be skewed by noise peaks. Since the difference in rest-frame UV and CO(3-2) positions are $<1$\,px for all sources, this has minimal effect on our analyses.

This process results in a stacked data cube, but we must also consider the effective PSF of the stack. To do this, we perform a parallel 2-D stack of the synthesized beam of each input data cube. Specifically, we examine the synthesized beam in each channel of an input cube, take the median of these beams, and place this median beam on the same spatial grid as the cube stack (i.e., $20''\times20''$, with $0.2''$/pixel). Each median beam is given the same weight as its corresponding data cube. Using a similar equation as above we find the 2-D PSF of the stack:
\begin{equation}\label{psfeq}
PSF_{stack}(x_i,y_j)=\frac{\Sigma_{n=1}^{N}PSF_{n}(x_i,y_j)w_n}{\Sigma_{n=1}^{N}w_n}
\end{equation}
This stacked PSF is then fit with a 2-D Gaussian, and the best-fit major axis, minor axis, and position angle are set as the PSF (i.e., synthesized beam) for each channel of the stacked cube.

We use each of the four weighting schemes to create a separate stacked cube (CASA immoments). These cubes are then collapsed over five velocity ranges ($\pm100\times[1,2,3,4,5]$\,km\,s$^{-1}$), resulting in 20 moment zero maps (see left panels of Figure \ref{bigbig}).

Note that we do not attempt to stack sources undetected in CO(3-2) emission. This is due to large differences between previously determined $z_{spec}$ values and the redshift of CO emission of SUPER sources (i.e., up to $\sim1500$\,km\,s$^{-1}$, see Table \ref{basictab}). Since the true systemic redshift of CO emission is not known for these sources, their CO emission could be significantly offset in velocity space.

\subsection{Radial Profile Extraction}\label{RPE}
\subsubsection{Method}
Since our goal is to search for low-level, extended CO emission (tracing a molecular gas reservoir), we extract radial brightness profiles of each moment zero map. We begin by finding the brightness-weighted spatial center by fitting a 2-D Gaussian to this emission. We then examine the emission in circular bins of width 1\,pixel centered on this spatial position. The values within each radial bin are collected, and the mean value in each bin is calculated. 

In order to determine the uncertainty on the mean in a bin, we take the maximum value of two values: the standard deviation in the bin and a beam-dependent RMS noise level (see Appendix \ref{uncanpp} for full details). 

Both the mean values and uncertainty are then normalized by dividing by the maximum value of the mean values. These normalized values are shown as the brown errorbars in the right panels of Figure \ref{bigbig}. The radial profile of the effective synthesized beam (shown as the cyan line in the right panels of Figure \ref{bigbig}) is derived in a similar fashion. 

In order to determine which results are physically significant, we consider the hypothetical case where the flux values contained within a radial bin are all distributed around 0, with a standard deviation equal to RMS$_{\mathrm{M0}}$. This is the ideal case of pure noise, and in this case the mean value would be 0. Considering an alternative case where the mean value is a positive value, the significance is more difficult to determine. This is because the emission is not necessarily circular, due to convolution with the beam, noise peaks, and a possibly complex intrinsic source morphology. Indeed, previous works (e.g., \citealt{fuji19,fuji20,herr21}) did not include a significance limit on their radial brightness profiles (although \citealt{gino20} presented a Poisson noise level). Since it is not possible to state a definite limit of significance in these radial profiles, we adopt the case where the mean value in a bin is equal to 0.5$\times$RMS$_{\mathrm{M0}}$ as an illustrative limit. Radial bins with mean values below this limit may still be significant, but as seen in Figure \ref{bigbig}, this is rarely the case. In addition, this limit allows us to see the peak significance of the detection (i.e., a stronger detection will result in a lower grey zone). Note that this choice is for data presentation only, and does not affect the analysis in the following sections.

When extracting the radial brightness profile of the moment zero map, the average S/N of each annulus decreases with increasing radius, so each profile features a point where noise dominates the signal ($\mathrm{r_{max}}$). Since the average intensity of annuli at this radius and beyond are not physically meaningful, we wish to exclude them from plots and further analysis. To determine this $\mathrm{r_{max}}$, we determine the smallest bin that features a negative intensity (i.e., $r_n$), and set $\mathrm{r_{max}}$ to the radius of the previous bin (i.e., $r_{n-1}$) . As seen in Figure \ref{bigbig}, this results in radial profiles with different maximum radii.

\subsubsection{Initial Results}\label{IR}
\begin{landscape}
\begin{figure}
    \centering
    \includegraphics[width=\linewidth]{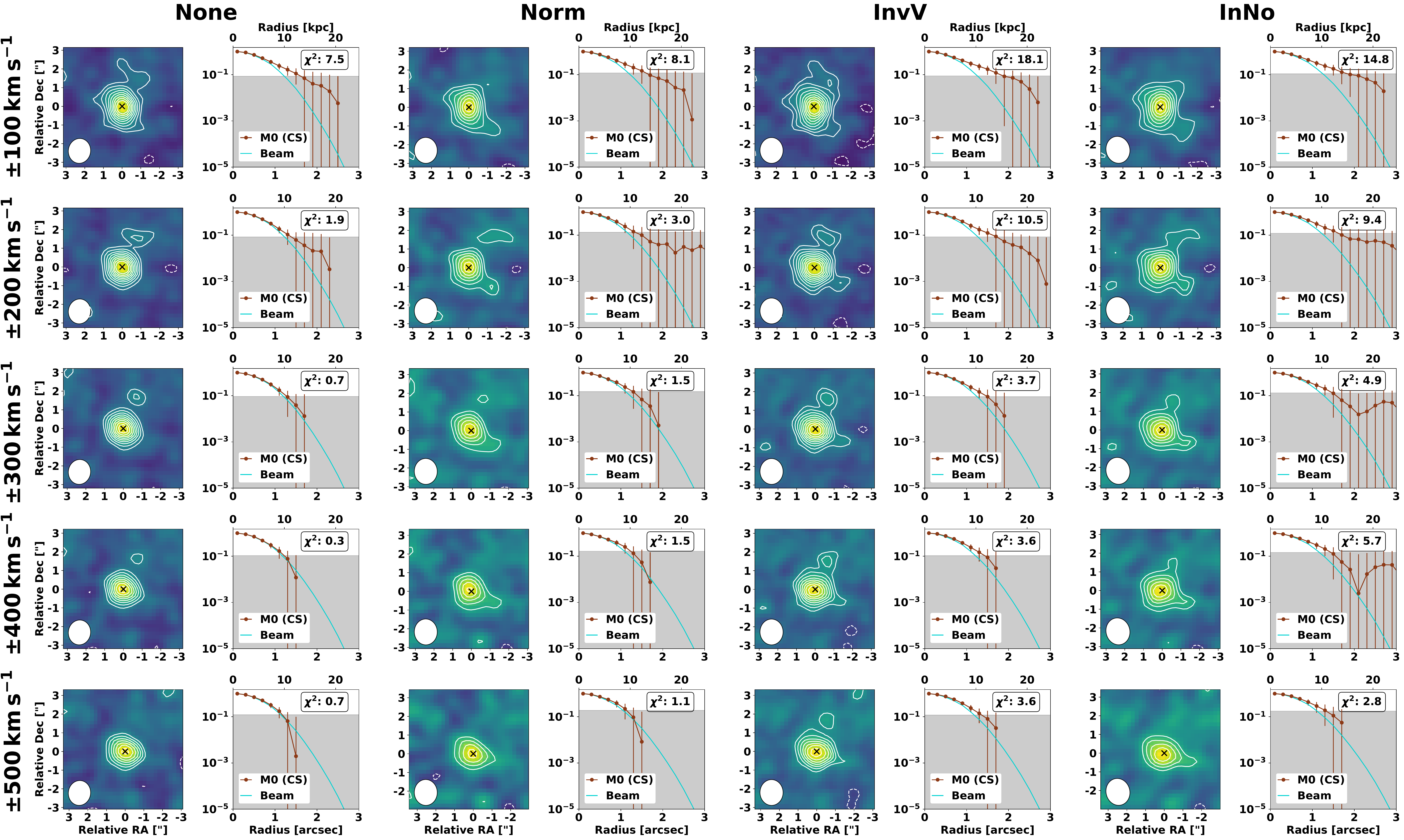}
    \caption{Moment zero maps (left) and radial brightness profiles (right) for each weighting technique (columns) and integrated velocity range (rows). The collapsed emission is shown with contours $\pm(2,3,4,\dots)\sigma$. The effective restoring beam is shown by the white ellipse to the lower left. For the right panels, the normalized radial profile of the moment 0 emission is shown in brown, with $1\sigma$ error bars, while that of the PSF is shown in cyan. The shaded region depicts a normalized mean intensity of $<0.5\times$(RMS noise level of the moment 0 map). The $\chi^2$ value representing how well the PSF profile matches the observed profile are presented in the upper right of each right panel. The legend label `M0 (CS)' stands for `Moment 0, Cube Stack', to differentiate these profiles from the Moment 0 stacking analysis presented in Section \ref{stmom}.
    }
    \label{bigbig}
\end{figure}
\end{landscape}

Before examining the radial profiles in the right panels of Figure \ref{bigbig} in a quantitative manner, we may examine their general appearance. If the emission is purely point-like, the brown and cyan profiles would line up exactly. On the other hand, an extended component smaller than the maximum recoverable scale (MRS) of the instrument will appear as a constant value (e.g., the grey uncertainty region) while pure noise or a diffuse component larger than the MRS would result in a value of 0 and would not be visible on these plots (i.e., log$_{10}$(x) approaches $-\infty$ as x approaches 0). If the source is resolved, then the slope of the brown line will be slightly shallower than the cyan. But perhaps the most interesting case is if the source may be decomposed into a central source and an extended component. In this case, we expect the extracted radial profile to show non-Gaussian behaviour due to the presence of multiple components.

For moment zero maps created with the same velocity range but different weighting schemes (i.e., plots in the same row), the spatial distribution of emission shows only slight variations between weighting schemes. This highlights the small intrinsic diversity of the sample. Since the sample of CO(3-2) data cubes we included in these stacks were all detected in line emission, it is not surprising that all 20 moment 0 maps show strong central detections. While most are in agreement with the PSF, some show significant extension. Because of this, we proceed by characterizing the relative contributions of the central and extended components.

%Updated FEB25
\pagebreak

To examine how these radial profiles deviate from the point source response, we examine the uncertainty-weighted chi-squared:
\begin{equation}\label{chi2}
\chi^2=\sum_{i=1}^{N}\left(\frac{D_i-M_i}{\delta D_i}\right)^2
\end{equation}
where $D_i$ is the observed normalized mean intensity for a radial bin, $\delta D_i$ is the associated uncertainty, and $M_i$ is an associated model value. Here, we treat the radial profile of the PSF as the `model' value and calculate the $\chi^2$ between the observed radial profile and the PSF. These values are shown in the upper right of each panel in Figure \ref{bigbig}.

As seen in this Figure, the radial profile of the `InvV'-weighted data cube over $\pm100$\,km\,s$^{-1}$ presents the largest $\chi^2$ value and thus deviates from a point-like source most strongly. The corresponding moment 0 map is clearly extended, and the radial profile shows complex behavior which suggests the presence of a second component. Since the average FWHM$_{\mathrm{CO}}$ of the stacked sources is $336\pm199$\,km\,s$^{-1}$ (see table B.1 of \citealt{circ21}), this represents integrating over the low-velocity, high-S/N emission near the line peak.

Since the moment zero map of this data cube stack shows the strongest deviation from a point-like source, it represents the best case for the existence of an extended component. While it is possible that a stronger deviation could be found using a different velocity bin (e.g., $\pm150$\,km\,s$^{-1}$), we will proceed with this image. In the next subsection, we explore whether this radial profile is better explained by a single resolved source or a central source with a low-level extended component. 

\subsection{Radial Profile Fitting}\label{HALOMOD}
\subsubsection{Methods}
Here, we explore whether the radial brightness profile of the moment zero image of the SUPER CO(3-2) stacked emission is better represented by a single resolved source or the combination of a central source and a diffuse component. 

While it would be possible to assume a general elliptical Gaussian (or 2-D S{\'e}rsic profile; \citealt{sers63}) and determine the intrinsic axis ratio and intrinsic position angle of the moment 0 map, we simply wish to search for the existence of diffuse emission, and thus proceed with a simpler 2-D circular Gaussian model. In addition, the stacking procedure averages these non-axisymmetric features.

To begin, we create a 2-D circular Gaussian model with arbitrary amplitude and set width (HWHM$_{\mathrm{G1}}$). This model is convolved with the representative PSF of a moment 0 map, and a radial profile of the convolved image is extracted. After normalizing the central value of this radial profile to unity, we compare the radial profiles of the moment 0 map and the convolved model. By using the Bayesian inference code MultiNest (\citealt{fero09}) and its python wrapper (PyMultiNest; \citealt{buch14}), we find the best-fit intrinsic FWHM of the model so that the difference between the radial profiles is minimized.

In this case, we only have one variable: the intrinsic width of the model source. We wish to explore a range of intrinsic widths, so we fit for log$_{10}($HWHM$_{\mathrm{G1}})$ (where the HWHM is in units of arcseconds) and set the prior to a uniform distribution between [-2,0.5], corresponding to angular scales [$0.01'',\sim3''$]. We adopt the likelihood function of \citet{niko09}:
\begin{equation}\label{nikoeq}
log_{10} L(\theta) = -\frac{1}{2} \sum_i \left[ \left[\frac{D_i-M_i}{\delta D_i}\right]^2 + log_{10} \left(2\pi \delta D_i^2	\right)\right]
\end{equation}
where each parameter is the same as in equation \ref{chi2}.

To test whether this radial profile could be explained by two components, we expand the initial model to include three parameters: the intrinsic width of a central source (log$_{10}($HWHM$_{\mathrm{G1}})$), the width of a more diffuse component (log$_{10}($HWHM$_{\mathrm{G2}})$), and the relative amplitudes of these two sources  ($\mathrm{log_{10}(f_{12})}$). The prior distributions for the two log$_{10}$(HWHM) variables are set to uniform distributions: [-2,0] (or [$0.01'',1''$]) for log$_{10}($HWHM$_{\mathrm{G1}})$ and [0,0.5] (or [$1'',\sim3''$]) for log$_{10}($HWHM$_{\mathrm{G2}})$. We adopt a uniform distribution between [-3.5,0) for $\mathrm{log_{10}(f_{12})}$.

\subsubsection{Results}
\begin{figure}
\centering
\includegraphics[trim=11.6cm 0 0 0, clip, width=0.5\textwidth]{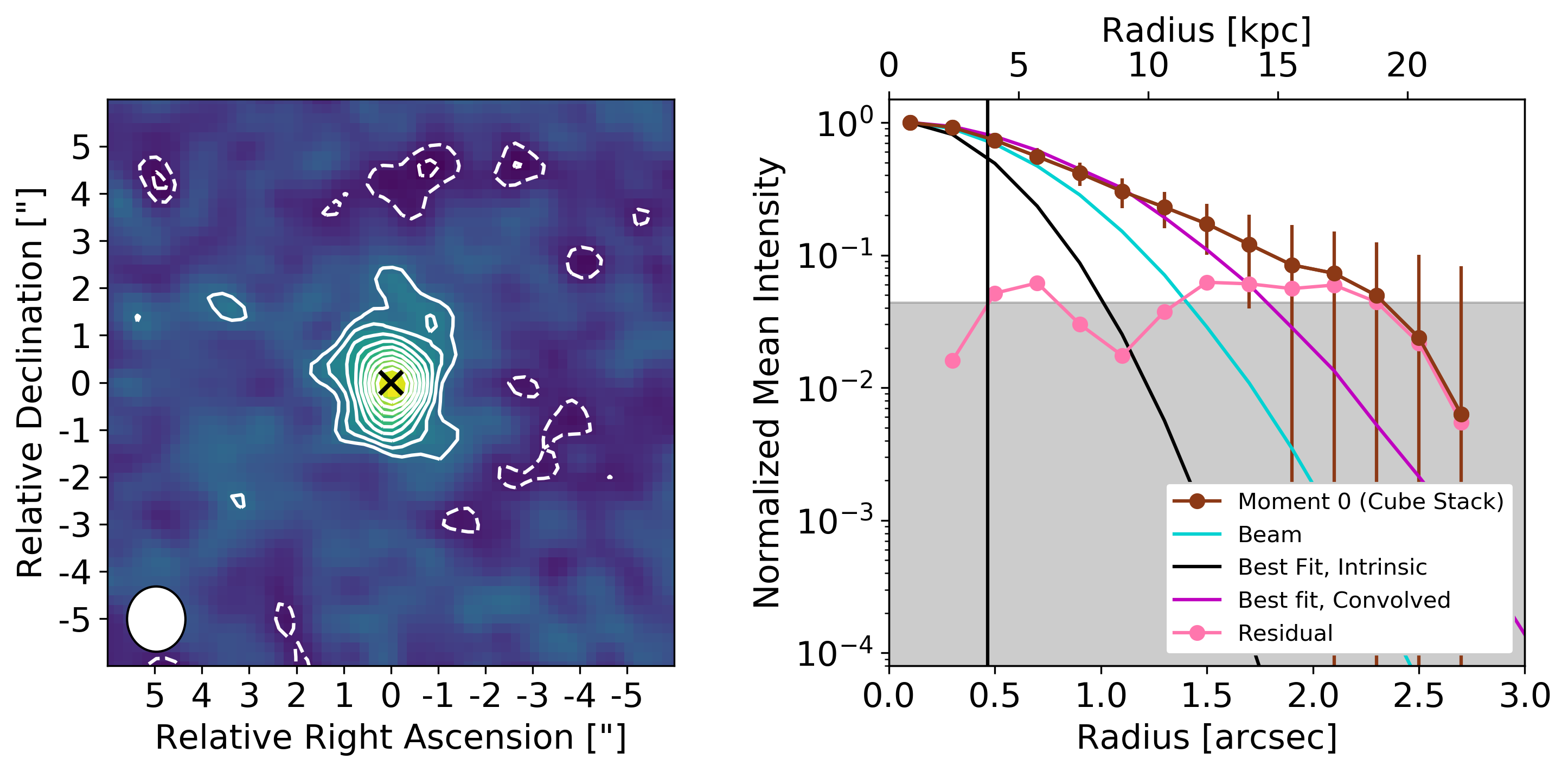}
\includegraphics[trim=11.6cm 0 0 0, clip, width=0.5\textwidth]{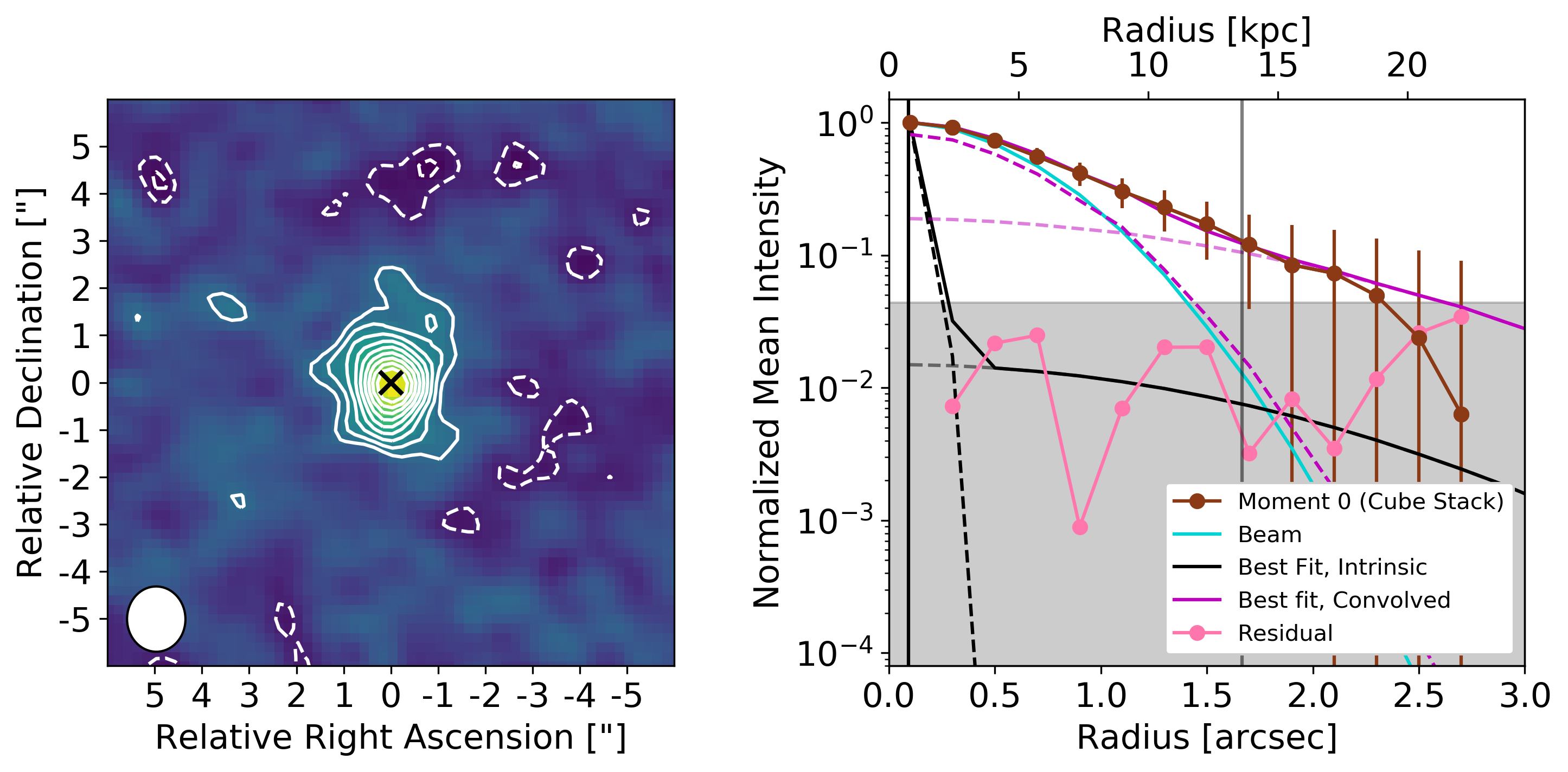}
\caption{Results of fitting single-Gaussian (top) and two-Gaussian (bottom) models to the $\pm100$\,km\,s$^{-1}$ moment 0 map of the InNo-weighted ALMA SUPER CO(3-2) stacked cube  (see Section \ref{HALOMOD}). In each panel, the brown and cyan lines show the normalized mean radial profiles of the moment 0 map and beam, while the shaded region depicts $<0.5\times$(RMS noise level of the moment 0 map). For the one-Gaussian model (top panel), we present the best-fit model radial profile (magenta) and intrinsic (i.e., unconvolved) profile (black solid curves), as well as the intrinsic HWHM value of the best-fit model (vertical black line) . For the two-Gaussian model (lower panel), the intrinsic and convolved components of the best-fit model are shown by dashed black and magneta lines, respectively. The vertical lines show the best-fit intrinsic HWHMs of each component. Each profile is normalized to its maximum value.}
\label{PMN_C}
\end{figure}

The results of fitting the one-component and two-component models to the $\pm100$\,km\,s$^{-1}$ moment 0 map of the InNo-weighted ALMA SUPER CO(3-2) stacked cube are shown in the top and bottom panels of Figure \ref{PMN_C}, respectively. The radial profiles of the moment 0 map, beam, and uncertainty are the same as in Figure \ref{bigbig}. However, we now denote the best-fit intrinsic HWHM by a solid black vertical line, and show its convolved radial profile with a magenta line. The residual between the model and data is shown by a green line. The returned best-fit parameters are listed in Table \ref{pmntable}, while the corner plots of each model are presented in Appendix \ref{corner}. 

%We note that each of the radial brightness profiles presented in this work are normalized to their brightest value. While the intrinsic profile (black line) features a brighter core than the convolved profile (magenta line) in the top panel of Figure \ref{PMN_C}, the values of their smallest-radius bin are set to be equal. The components of the two-Gaussian fit in the lower panel are each normalized by the maximum value of their sum. Thus, the intrinsic weak component (black dashed line with low slope) does not `gain flux' when convolved despite the fact that the corresponding dashed magenta line is higher. The convolved model simply has a lower maximum amplitude and thus a smaller normalization.

While the single-component fit captures the $r<10$\,kpc behaviour quite well, there are slight residuals above the 0.5$\times$RMS level at both small and large radius (Figure \ref{PMN_C}). By adding a second component, the best-fit model radial profile is a better match, and the residuals at large radius are much reduced. These results argue that the stacked emission is best explained by a central source with a low-level, diffuse component. However, this interpretation is primarily qualitative, so we now turn to a quantitative analysis of the relative goodness of fit for each model.

\subsubsection{Goodness of Fit}
This goodness of fit may be made quantitative in several ways. The most simple is to use an uncertainty-weighted $\chi^2$ (Equation \ref{chi2}). The two-component model is preferred, as $\chi^2$ is lower by a factor of $\sim7$ (see Table \ref{pmntable}). We note that the two-component model has a $\chi^2$ value that is sub-unity, which usually indicates overfitting. However, this is likely due to our conservative uncertainty limits (see Appendix \ref{uncanpp}).

While this value reflects how well the data and model agree, it does not encapsulate the complexity of the respective models. Indeed, when comparing models with different numbers of fit variables, it is crucial to take degrees of freedom into account to avoid overfitting. Because of this, we also use the reduced $\chi^2$ (e.g, \citealt{webb21}):
\begin{equation}
\chi_{red}^2=\frac{\chi^2}{N_{data}-N_{variables}}
\end{equation}
where $N_{data}$ is the number of fit data points and $N_{variables}$ is the number of free parameters in the model. This value inform us if the improvement in goodness of fit in more complex models is significant. Again, the two-component model is preferred, suggesting that we are not overfitting.

We may also use the Bayesian evidence of each fit (as output by PyMultiNest) to judge the quality of each fit. MultiNest operates on the princple of Bayes' theorem (e.g., \citealt{fero09}):
\begin{equation}
Pr(\theta|D,H)=\frac{Pr(D|\theta,H)Pr(\theta|H)}{Pr(D,H)}
\end{equation}
where $Pr(\theta|D,H)$ is the posterior probability distribution for a set of parameters $\theta$, $Pr(D|\theta,H)$ is the likelihood, $Pr(\theta|H)$ is the prior, and $Pr(D,H)$ is the Bayesian evidence (also written as $Z$). When considering two models, we may use their relative Bayesian evidences to calculate the Bayes factor:
\begin{equation}
K=\frac{Z_1}{Z_2}
\end{equation}
where we assume that neither model is preferred \textit{a priori}. The resulting value of $K$ informs us of which model is favored: $K\sim1-10$ is weak evidence towards model 1, $K\sim10-30$ is strong evidence towards model 1, and $K>30-100$ is very strong evidence for model 1 \citep{jeff61}. The resulting Bayesian evidence values are listed in Table \ref{pmntable}. We find that the Two Gaussian model is strongly favored ($K\sim27$). \footnote{Due to the low number of data points, we are not able to use the Bayesian Information Criterion (BIC) or Akaike Information Criterion (AIC; e.g., \citealt{lidd07,conc19}).}

These three tests agree that the two-component model is preferred over an unresolved source or the single-component model, with little evidence for overfitting and slight evidence for overestimated uncertainties. This best fit model includes a compact, barely resolved source with an amplitude of unity and HWHM=$0.09^{+0.15}_{-0.06}=0.76^{+1.19}_{-0.46}$\,kpc, and a more diffuse component with an amplitude of $0.005^{+0.004}_{-0.002}$ and HWHM=$1.66^{+0.58}_{-0.43}"=13.49^{+4.71}_{-3.49}$\,kpc. Note that these values were fit in log-space, and so the listed uncertainties are $1\sigma$ errors on the logarithm of each parameter. Each value is PSF deconvolved. In the next Section, we discuss the implications of this result.

\begin{table*}
\centering
\begin{tabular}{cc|ccc|ccc}
                &	            & log$_{10}($HWHM$_{\mathrm{G1}}/[$UNIT$])$ & log$_{10}($HWHM$_{\mathrm{G2}}/[$UNIT$])$ & log$_{10}($f$_{\mathrm{12}})$	& $\chi^2$ & $\chi_{red}^2$ & ln(Z)\\
Stack Type      & Model	        & [$"$], [kpc]	                            & [$"$], [kpc]	                            &		                        &		   &   	            &      \\ \hline

Cube Stack      & PSF           & $\times$                                  & $\times$                                  & $\times$                      & 18.08    & $\times$       & $\times$ \\                                                
                & One Gaussian	& $-0.33\pm0.06, 0.59\pm0.06$	            & $\times$	                                & $\times$	                    & 4.14     & 0.32           & $40.14\pm0.08$\\ 
                & Two Gaussian  & $-1.03\pm0.41, -0.12\pm0.41$              & $0.22\pm0.13, 1.13\pm0.13$                & $-2.27\pm0.23$                & 0.58     & 0.05           & $42.47\pm0.07$\\ \hline

Moment 0 Stack  & PSF           & $\times$                                  & $\times$                                  & $\times$                      & 7.42     & $\times$       & $\times$ \\    
                & One Gaussian	& $-1.05\pm0.41, -0.14\pm0.41$	            & $\times$	                                & $\times$	                    & 5.80     & 0.41           & $50.50\pm0.04$\\
                & Two Gaussian  & $-0.95\pm0.25, -0.04\pm0.25$              & $0.28\pm0.14, 1.19\pm0.14$                & $-2.88\pm0.36$                & 3.16     & 0.26           & $50.83\pm0.07$\\ \hline

Continuum Stack & PSF           & $\times$                                  & $\times$                                  & $\times$                      & 5.16     & $\times$       & $\times$ \\    
                & One Gaussian	& $-1.18\pm0.34, -0.26\pm0.34$	            & $\times$	                                & $\times$	                    & 6.50     & 0.34           & $70.24\pm0.04$\\
                & Two Gaussian  & $-0.88\pm0.12, 0.04\pm0.12$               & $0.28\pm0.14, 1.19\pm0.14$                & $-3.08\pm0.26$                & 5.54     & 0.33           & $69.63\pm0.09$\\ \hline 

\end{tabular}
\caption{Best-fit values for a point source (PSF), one-Gaussian model, and two-Gaussian model applied to SUPER galaxies: the $\pm100$\,km\,s$^{-1}$ CO(3-2) moment 0 map of the InNo-weighted stacked cube (`Cube Stack', see Section \ref{HALOMOD}), the InNo-weighted stacked CO(3-2) moment 0 map (`Moment 0 Stack', Section \ref{stmom}), and the InNo-weighted stacked FIR continuum map (`Continuum Stack', see Section \ref{stcon}). We also note the $\chi^2$, $\chi_{red}^2$, and Bayesian evidence for each.}
\label{pmntable}
\end{table*}

\section{Discussion}\label{disc}

\subsection{Additional Stacking Analyses}
The primary analysis of this work has been based on a three-dimensional stack of seven ALMA CO(3-2) images of $z\sim2-2.5$ AGN host galaxies from the SUPER sample. In addition, we may examine multiple 2-D stacks of these sources. In this subsection, we detail stacking analyses of HST data (Section \ref{hstsect}), CO(3-2) moment 0 maps (Section \ref{stmom}), and rest-frame FIR continuum maps (Section \ref{stcon}).

\subsubsection{HST stacking}\label{hstsect}
If the extended emission found in the previous Section is tied to the inner CGM, its radial brightness distribution should differ from that of stars and the gas around them (i.e., the interstellar medium; ISM). Here, we analyze rest-frame UV images (tracing the young stellar distribution) of SUPER galaxies in order to test this hypothesis.

The majority of the sample examined by \citet{circ21} (21/28 sources) lie in the COSMOS field, and thus benefit from a host of multiwavelength observations (\citealt{scov07a,scov07b})\footnote{The remaining seven sources are from the XMM-XXL north field, which lacks suitable HST observations.}. We choose to analyze the HST/ACS F814W (i-band) data for each of these sources. As the reddest broad optical filter for ACS, F814W reveals the morphology of the young stellar population (e.g., \citealt{pope11}) with higher sensitivity than other filters (e.g., F775W; \citealt{koek07}).

Each i-band image was downloaded from the IPAC COSMOS server\footnote{\url{https://irsa.ipac.caltech.edu/data/COSMOS/index_cutouts.html}}, using the central positions given by \citet{circ21} as the centre of a $15''\times15''$ cutout. Due to the AGN-nature of this sample, all 21 sources feature strong central i-band detections. These images have cell sizes of $0.03''$, and we assume a circular PSF of FWHM $0.095''$ \citep{koek07}. We exclude the three sources observed to have close companions or disturbed morphologies (CID\_971, CID\_1215, CID\_1253; see Section \ref{obsdat}).

While the CO data were stacked in three dimensions, the i-band images must be stacked in two dimensions. Thus, we use equation \ref{psfeq} with the `inverse variance' weighting scheme (InvV). We fit a 2-D Gaussian to the central emission of each input image and set the centre of each galaxy as the best-fit centroid, in order to account for offsets between gas and stellar emission. A radial profile is then taken, using annuli that are 2\,pixels wide.

\begin{figure*}
    \centering
    \includegraphics[width=0.6\textwidth]{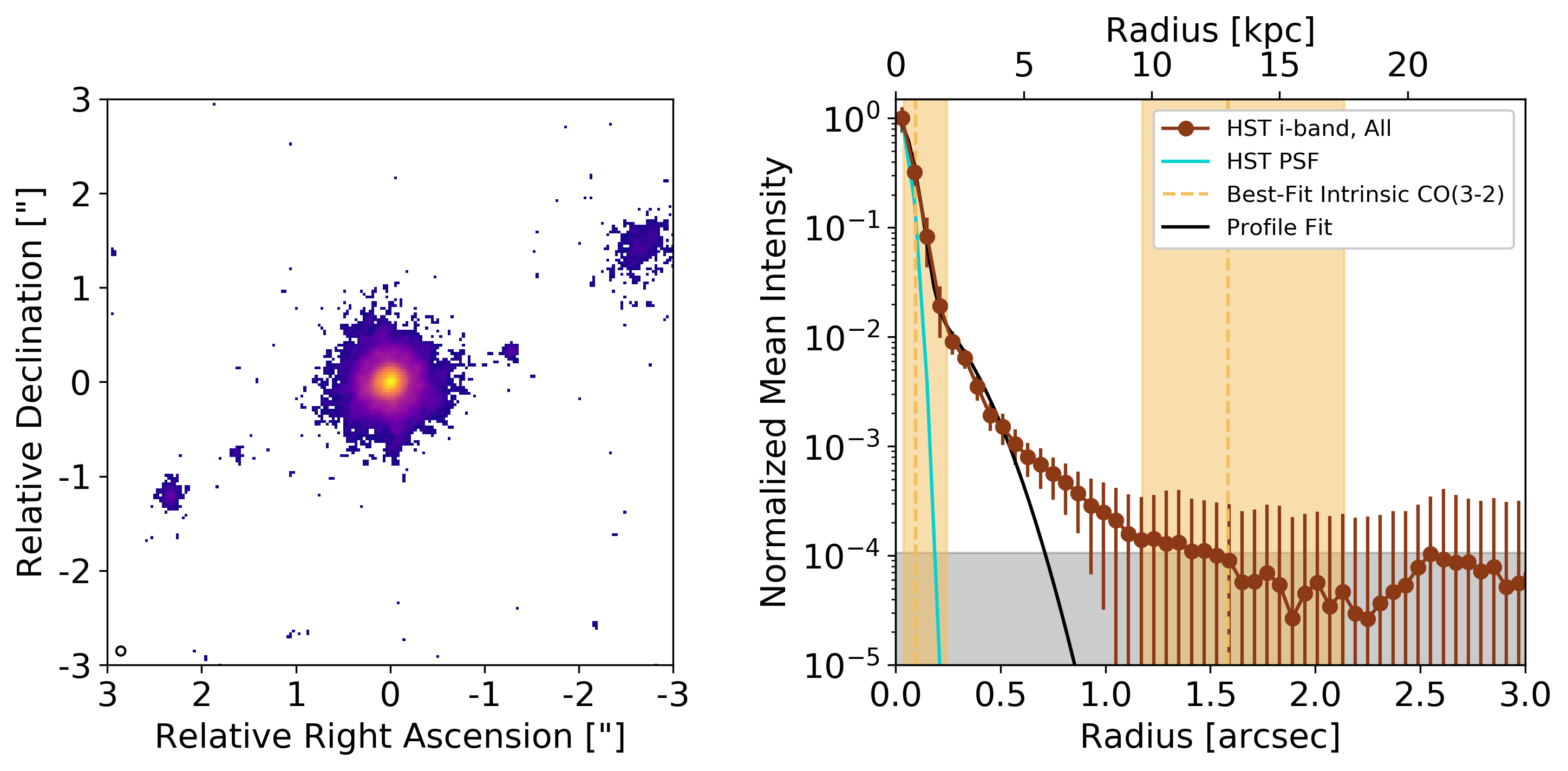}
    \caption{Intensity map (left) and radial brightness profiles (right) for the stacked image of HST/ACS i-band emission using 21 SUPER galaxies. The PSF ($0.095''$) is shown by the ellipse to the lower left. In the right panel, the normalized radial profile of the intensity map is shown in brown with error bars (showing standard deviation within annulus), while that of the PSF is shown in cyan. The shaded grey region depicts a normalized mean intensity of $<0.5\times$(RMS noise level of the intensity map). The two best-fit intrinsic radii of the CO(3-2) moment 0 map of the InNo-weighted stacked ALMA SUPER CO(3-2) cube (see Figure \ref{PMN_C}) are shown by dashed yellow (with $1\sigma$ errors) for reference. A two-Gaussian fit to the brightness profile is depicted in black.}
    \label{allhst}
\end{figure*}

The results of stacking all 21 HST/ACS i-band images and taking the radial brightness profile of the resulting image is shown in Figure \ref{allhst}. Since the PSF is much smaller than the ALMA observations (i.e., $0.095''$ rather than $\sim1''$), the radial profile is much less affected by convolution effects. The emission is clearly detected and resolved, with a clear divergence from what is expected from a simple single-component source. To illustrate this, we performed a two-component Gaussian fit to the observed brightness profile (using scipy \textlcsc{curve\_fit}). The inner emission ($r\lesssim0.25''$) is well captured by a strong, narrow component (HWHM$\sim0.2''$). However, the profile between $r\sim0.5-1.5''$ shows a much more shallow slope, which is poorly fit with a Gaussian (HWHM$\sim0.5''$). However, it may be better fit with a general S\'{e}rsic profile. 

If this profile is compared to the best-fit intrinsic HWHM values of the CO(3-2) moment 0 map (yellow lines and shaded uncertainty regions), we see that the significant emission barely reaches the scale of the larger CO HWHM, while the smaller CO HWHM agrees with the $<0.25''$ behaviour of the HST radial profile, which likely represents the host galaxy. Since the CO emission shows emission beyond $10$\,kpc, this suggests that the molecular gas extends beyond the stellar component.

%Updated FEB28

\begin{figure*}
    \centering
    \includegraphics[width=0.6\textwidth]{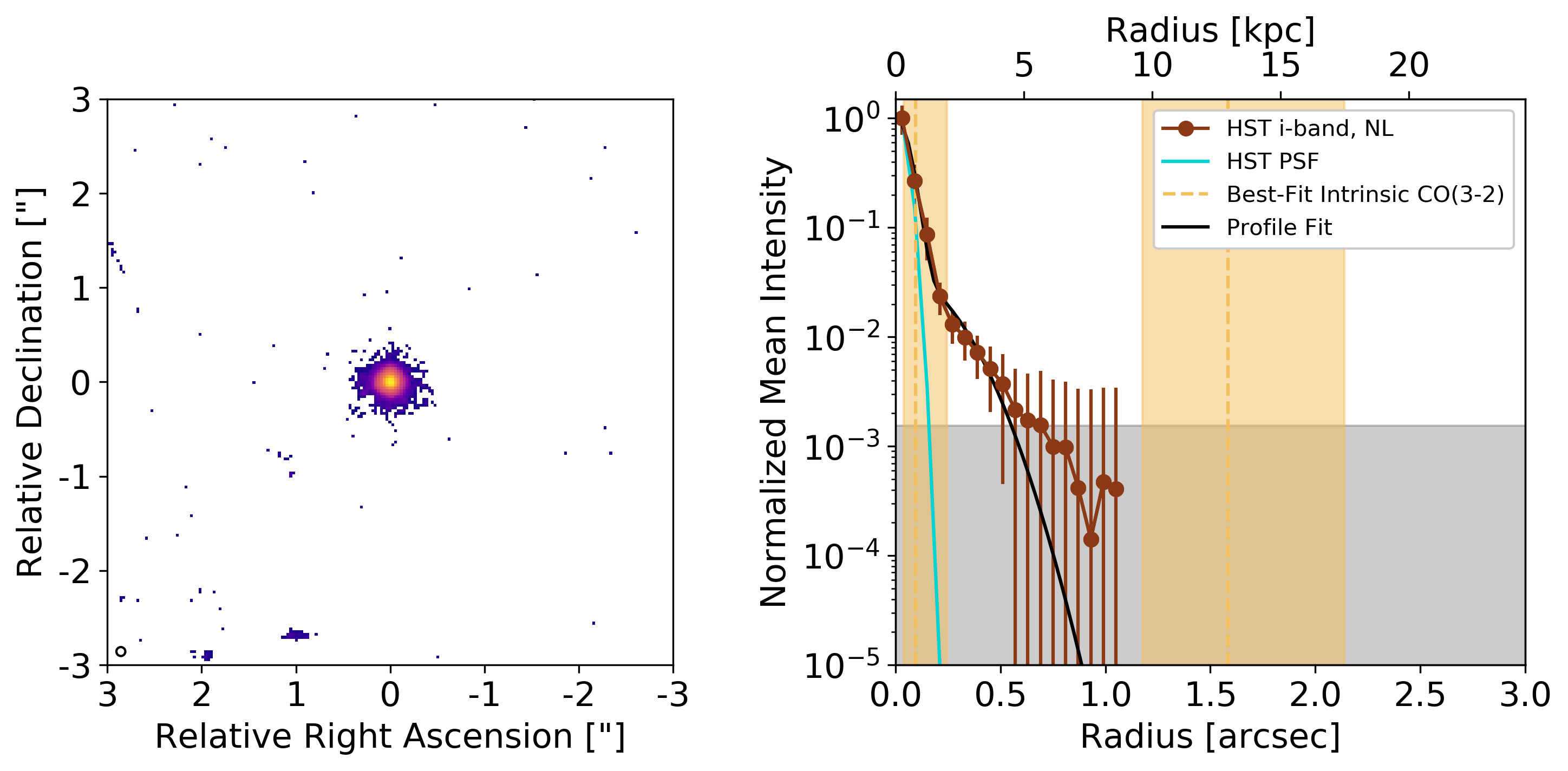}
    \includegraphics[width=0.6\textwidth]{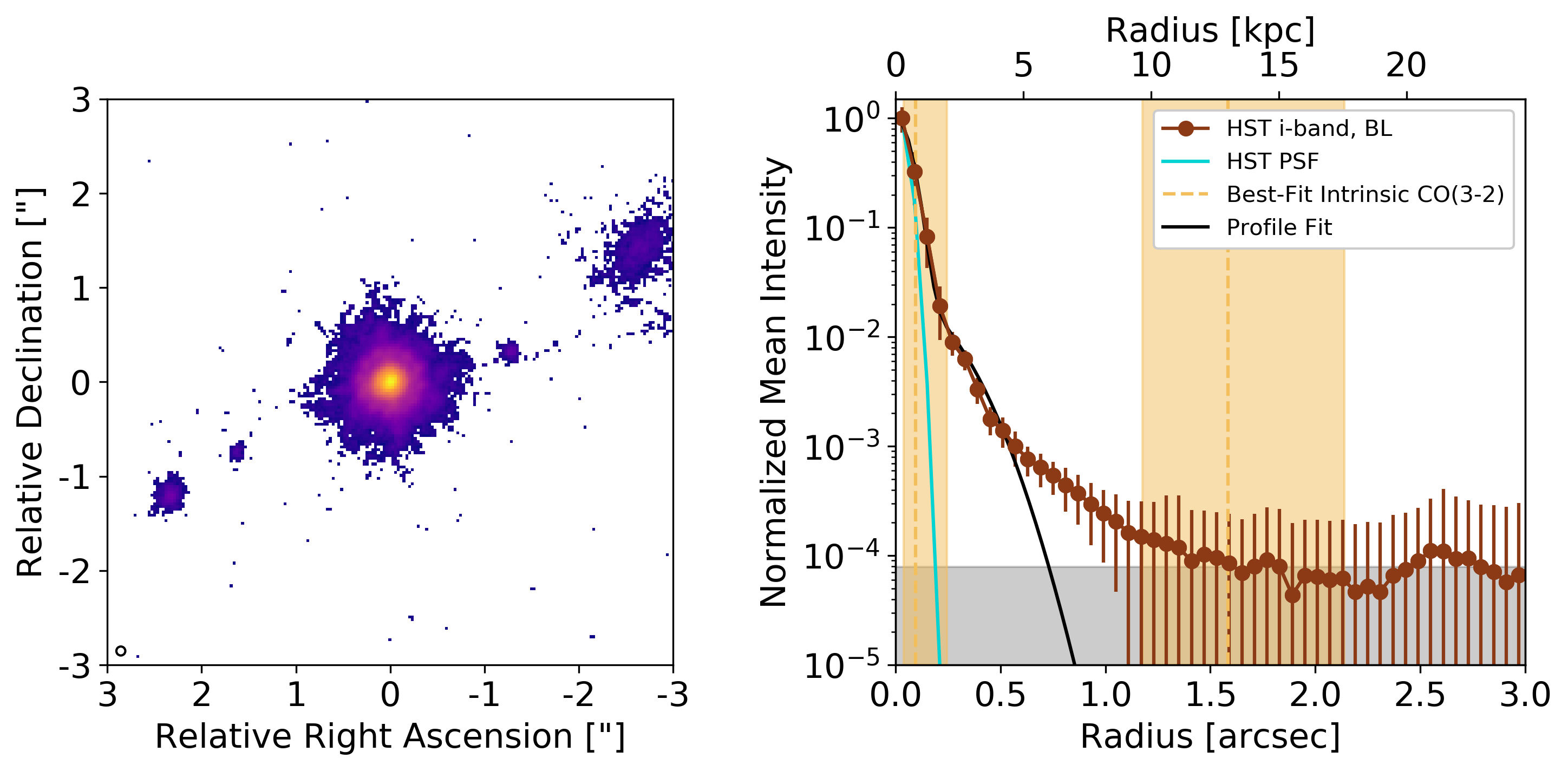}
    \caption{Intensity map (left) and radial brightness profiles (right) for stacked HST/ACS i-band images using 11 NL (top) and 10 BL (bottom) SUPER galaxies. The PSF ($0.095''$) is shown by the ellipse to the lower left. In the right panel, the normalized radial profile of the intensity map is shown in brown with $1\sigma$ error bars, while that of the PSF is shown in cyan. The grey shaded region depicts a normalized mean intensity of $<0.5\times$(RMS noise level of the intensity map). The two best-fit intrinsic radii of the CO(3-2) moment 0 map of the InNo-weighted stacked ALMA SUPER CO(3-2) cube (see Figure \ref{PMN_C}) are shown in yellow (with $1\sigma$ errors) for reference. A two-Gaussian fit to the brightness profile is depicted in black.}
    \label{somehst}
\end{figure*}

The sample of SUPER sources with HST/ACS i-band photometry includes both NL and BL AGN, where BL AGN were classified by the presence of broad (FWHM$>10^3$\,km\,s$^{-1}$) emission lines in optical spectra. Since in the case of BL AGN the continuum image may be dominated by the nuclear AGN, it is important to test whether there are any differences between the two sub-sample and, in particular, whether the NL sample presents the same compact emission. Therefore, we may divide the 21 sources into these groups (11 NL and 10 BL sources). The results of stacking these groups and extracting radial profiles are shown in Figure \ref{somehst}. We may see that the NL stack has similar small-scale behaviour as the total stack, but with a lower S/N. The BL sample features stronger i-band detections, resulting in a much higher central S/N of the lower radial profile in Figure \ref{somehst}. 

Despite the differences between the two groups, each radial profile shows the same qualitative behaviour as the full sample: a steep decrease from the central radial bin to $\sim0.2''$ and a bump between $\sim0.2-0.5''$. While the initial decrease may easily be fit with a convolved Gaussian source, the outer radial profile quickly deviates from this simple fit. Since the ALMA data have much coarser resolution (i.e., $\sim1''$), we may not state whether this complex behaviour is due to stacking galaxies with a diversity of small-scale properties or the effect of outflows or complex source morphologies (e.g., disturbed disk, close-separation mergers). However, we may state that the stacked HST/ACS i-band data shows no evidence for significant extended emission beyond $\sim7.5$\,kpc, unlike the ALMA CO(3-2) data. This suggests that the gaseous reservoir traced by CO(3-2) is morphologically separate from the stellar distribution, and thus may trace the inner CGM rather than the ISM. On the other hand, the fact that the molecular gas has a greater extent than the young stellar population may simply indicate an inside-out growth of stellar mass (e.g., \citealt{fran19}). This gas would then be directly associated with the host galaxy, and would be used in future star formation.

%Updated FEB28

In order to compare these results more directly with those of the 3-D CO stacking analysis, we repeat this analysis using the four galaxies included in the 3-D stacking analysis that have i-band images. Since all of these are BL AGN, the resulting radial profile is nearly identical to the profile of the full BL sample shown in Figure \ref{somehst}, and is therefore not included. 

Finally, we follow a similar approach to past comparisons of ALMA and HST data (e.g., \citealt{fuji19}) and explore the effect of convolving the high-resolution HST data (PSF$\sim0.095''$) with the ALMA beam ($1.38\times1.24$ at $0.58^{\circ}$). The result of this convolution is shown in the left panel of Figure \ref{PMN_HST}. We also fit the radial profile of this convolved map, finding that a single Gaussian component ($r\sim2$\,kpc) is a better fit than the PSF, and a double Gaussian component ($r\sim1$\,kpc and $r\sim18$\,kpc) returns the best fit (based on $\chi^2$ and $\chi_{red}^2$ statistics).

At first glance, the fact that the best-fit model includes a spatially extended component that is comparable in size to that of the CO emission may imply that the gas and stellar component are well-linked, and that the earlier disagreements were simply a matter of resolution. In this light, the `extended' CO component may simply be a result of a complex compact gas morphology that is convolved with a large beam.

The second way to interpret this, however, is to first examine the intrinsic (i.e., deconvolved) stacked HST image (Figure \ref{allhst}). This emission is clearly extended past the PSF, but also features multiple discrete companions that are $\sim1-2''$ away from the central source. These may be true satellite galaxies, or low-redshift interlopers. These other sources will naturally cause an effect on the radial profile, and may be to blame for the deviation from a single resolved source.

In summary, our test of convolving the HST stacked image to the ALMA resolution reveals a radial brightness distribution is deviates from a single resolved source, but it is not clear if this is due to resolution effects or the presence of foreground galaxies.

\subsubsection{Moment 0 map stacking}\label{stmom}
The three-dimensional stacking of this work is dependent on a number of assumptions, including a lack of diversity between sources (see Section \ref{stas} for a more complete discussion). Indeed, \citet{circ21} found that the CO(3-2) profiles of the SUPER sample feature different linewidths. In order to account for this, we may first create a CO(3-2) moment zero map for each source, using all channels identified to contain line emission. As seen in Table \ref{m0size}, the emission in these maps is either unresolved or poorly resolved (i.e., each size estimate is within $3\sigma$ of zero).

These moment 0 maps may then be stacked in 2-D space, using equation \ref{psfeq}. We explore each of the four weighting schemes introduced in Section \ref{stamed}, finding that the `InvV' scheme results in the greatest divergence from a point source (as quantified by the $\chi^2$ comparison between the beam and data profiles).

To explore the presence or absence of an extended component, we repeat the analysis of Section \ref{HALOMOD} (i.e., using a one- and two-component model to fit for the intrinsic spatial scale of the emission), but now using the stacked moment zero map rather than a moment zero map created using a stacked data cube. 

\begin{figure*}
\includegraphics[width=0.6\textwidth]{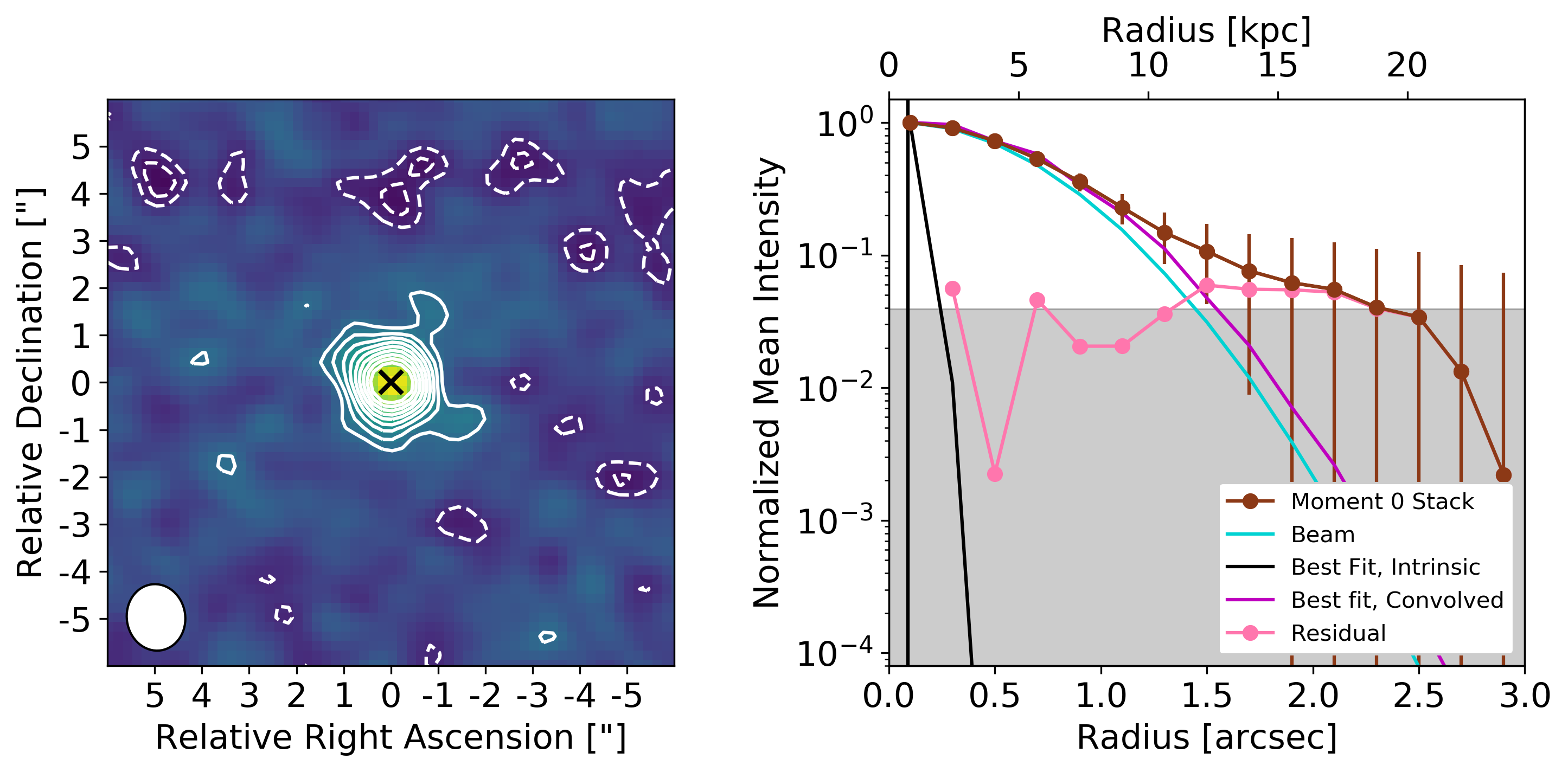}
\includegraphics[trim=11.6cm 0 0 0, clip, width=0.3156\textwidth]{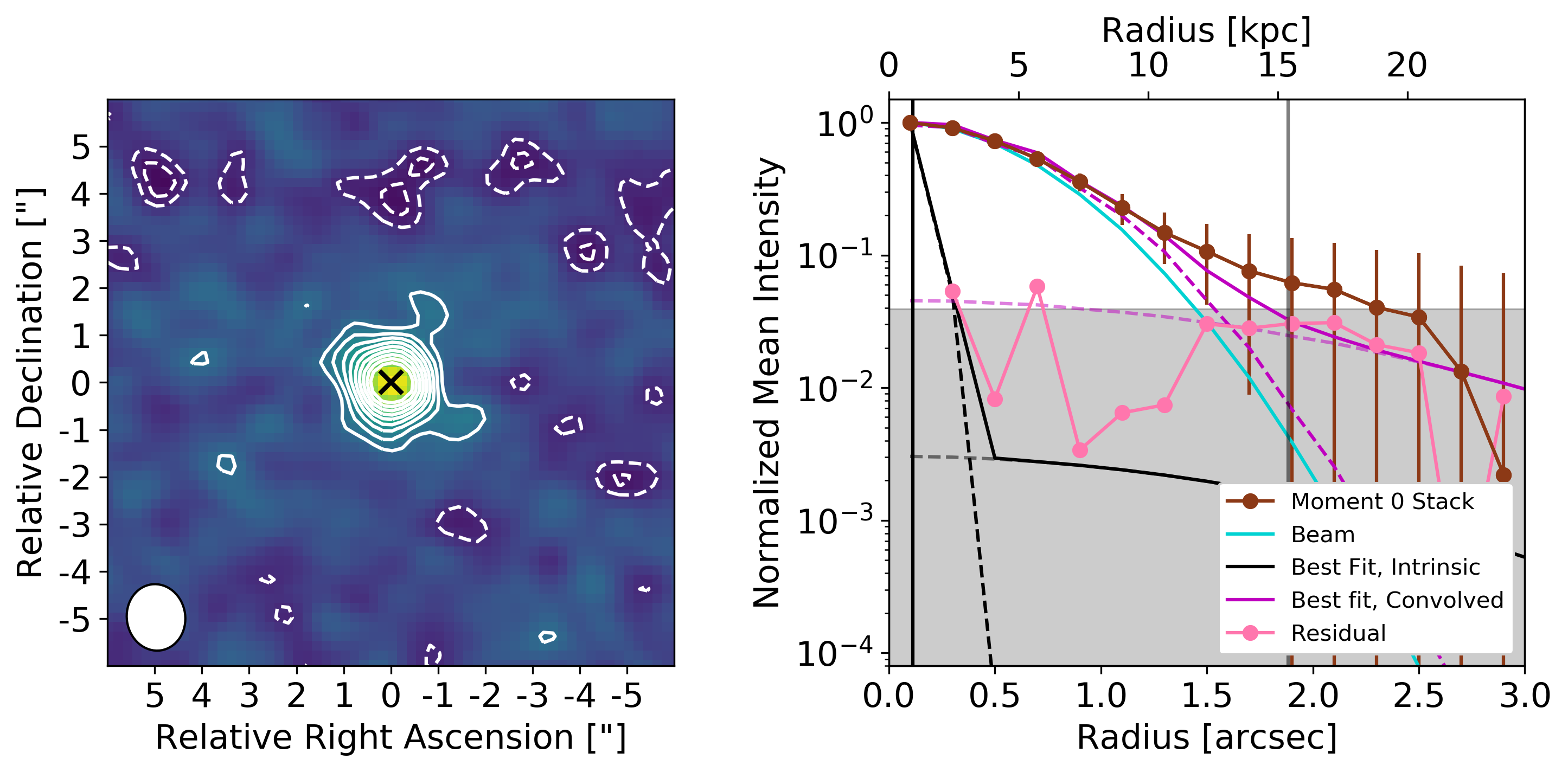}
\caption{Intensity map (left) and best-fit radial brightness profile for stacked CO(3-2) moment 0 maps using InvV-weighting and a one-component (centre) or two-component model (right). The collapsed emission is shown with contours $\pm(2,3,4,\dots)\sigma$. The synthesized beam is shown by the ellipse to the lower left. In the centre and right panels, the normalized radial profile of the intensity map is shown in brown with $1\sigma$ error bars, while that of the PSF is shown in cyan. The grey shaded region depicts a normalized mean intensity of $<0.5\times$(RMS noise level of the intensity map).}
\label{mom0stack_fig}
\end{figure*}

The resulting map, radial profile, and best-fit models are presented in Figure \ref{mom0stack_fig}, and the best-fit parameters and goodness of fit are listed in Table \ref{pmntable}. It is clear that this emission is best-fit by a two-Gaussian model, as quantified by a smaller $\chi^2$ and $\chi_{red}^2$. However, the Bayes factor does not show a preference for either model (Z$_2$/Z$_1\sim2$). The best-fit parameters of the Two Gaussian model are in agreement with the results of the `Cube Stack' analysis, to within $2\sigma$.

\begin{table}
\centering
\begin{tabular}{c|c}
Source & Size\\ \hline
CID\_166 & $(2.13\pm0.95)''\times(0.98\pm0.56)'', 18\pm34^{\circ}$\\
CID\_346 & $(0.71\pm0.28)''\times(0.54\pm0.21)'', 154\pm140^{\circ}$\\
CID\_451 & $(2.27\pm0.85)''\times(2.17\pm0.96)'', 137\pm124^{\circ}$\\
LID\_206 & $\lesssim0.7''$\\
X\_N\_44\_64 & $\lesssim0.6''$\\
X\_N\_53\_3 & $\lesssim0.6''$\\
X\_N\_81\_44 & $(2.47\pm0.87)''\times(0.30\pm0.45)'', 47\pm10^{\circ}$\\
\end{tabular}
\caption{Results of fitting 2-D Gaussian models to the CO(3-2) moment map of each source. Sizes are beam-deconvolved, while presented size limits are the geometric mean of the major and minor HWHMs of each synthesized beam (e.g., \citealt{miet17}).}
\label{m0size}
\end{table}

The different appearance of the radial profiles between these two approaches is due to the fact that the first approach (`Cube Stack') includes only the high-signal channels at $|v|\leq100$\,km\,s$^{-1}$, while the second approach (`Moment 0 Stack') includes all channels identified to contain line emission ($|v|\lesssim100-600$\,km\,s$^{-1}$, depending on source). Since a moment 0 map is proportional to the sum of the values in each pixel:
\begin{equation}
M_0(x,y)=\Delta_v\Sigma_i^NS(x,y,v_i)
\end{equation}
the inclusion of low-S/N data at high $|v|$ will wash out the low-amplitude extended signal. 

On the other hand, if we create and stack moment zero maps using a uniform range of $|v|\leq100$\,km\,s$^{-1}$, we recover a nearly identical image to the `Cube Stack' result. This is because both approaches combine data from a set of 3-D data cubes into a single image, but use different orders of operation.

\subsubsection{FIR Continuum stacking}\label{stcon}
We have now examined the distribution of CO(3-2) (tracing molecular gas) and HST/ACS i-band emission (tracing the stellar distribution). It is also possible to examine the rest-frame FIR continuum emission ($\nu_{rest}\sim870\,\mu$m). Other works found that in a number of high-redshift sources (e.g., $z\sim4-9$ SFGs detected in FIR continuum emission; \citealt{fuji20,fuda22}, a $z\sim6$ QSO host galaxy; \citealt{meye22}), the dust emission featured a slightly smaller spatial extent, but was comparable to that of the \cii emission.

There are four sources in our sample of 28 ALMA-observed sources that show signatures of mergers or close companions ($<10''$) in CO and/or continuum images (CID\_1215, X\_N\_6\_27, CID\_1253, and CID\_971) which we exclude. For the remaining 24 sources in our sample (see Table \ref{basictab}), we create continuum maps by using the CASA task \textlcsc{tclean} in `MFS' mode, selecting the line-free channels (see Section \ref{obsdat}). The RMS noise level in this `dirty' image is identified, and a `clean' image is made by cleaning down to $3\times$ this RMS noise level. 

Due to significant differences in $z_{spec}$ and $z_{CO}$, the CO line emission stacks (both 2-D and 3-D) only included CO-detected sources. This was done in order to maximize our S/N and avoid including a line-less spectrum. Unlike the CO stack, the continuum stack may include continuum-undetected sources, as there is no concern with redshifts. Each continuum image was made by excluding CO emission based on either $z_{CO}$ (if it was known) or $z_{spec}$. So for the CO-undetected sources, there may be low-level line emission that adds extra flux to  the continuum images. This is not a concern for this work, as any line emission would be quite weak (i.e., it is not significant enough to be detected in the cube). 

\begin{figure*}
\includegraphics[width=0.6\textwidth]{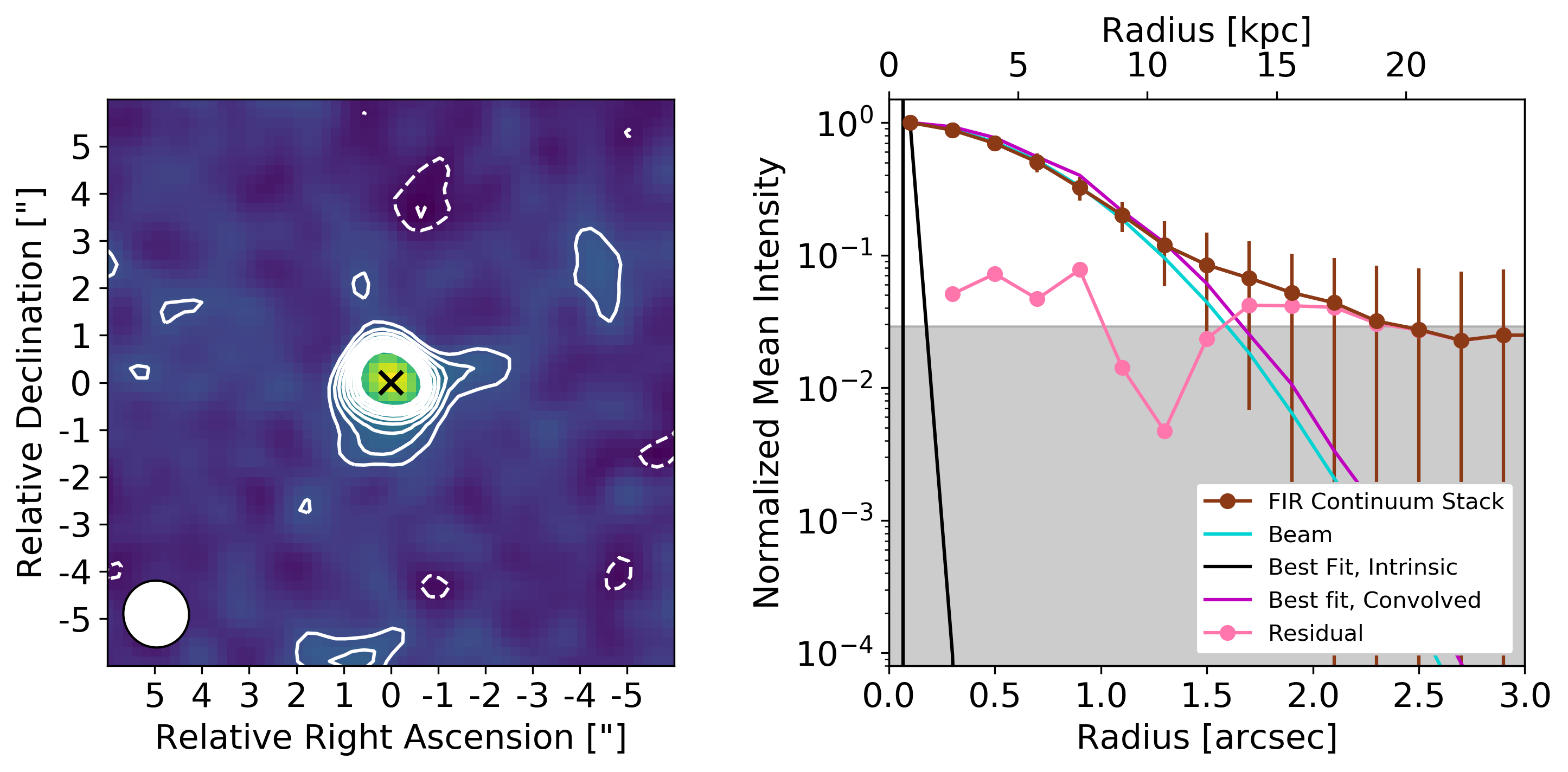}
\includegraphics[trim=11.6cm 0 0 0, clip, width=0.3156\textwidth]{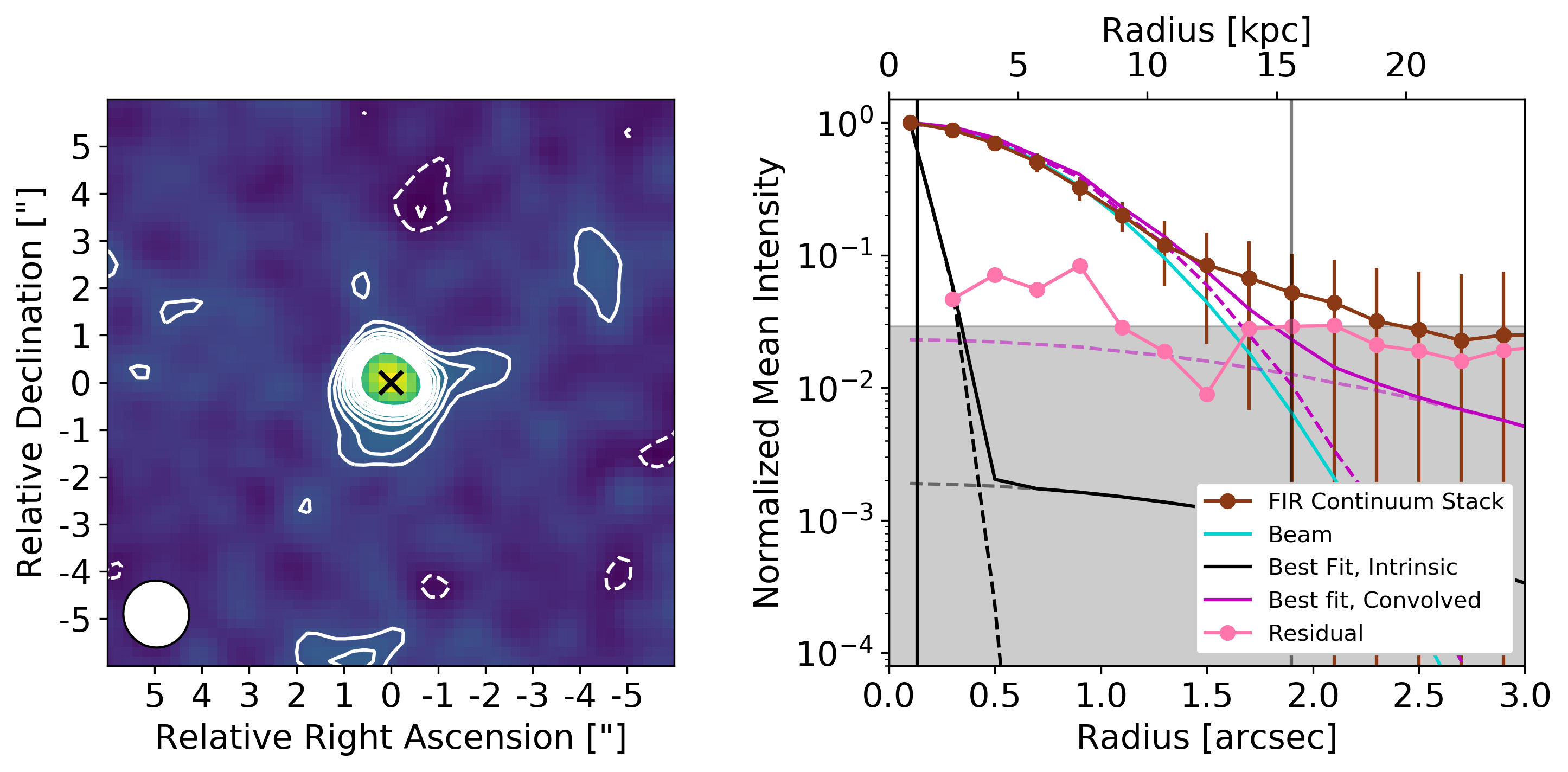}
\caption{Intensity map (left) and best-fit radial brightness profile for stacked rest-frame FIR continuum maps using InvV-weighting and a one-component (centre) or two-component model (right). The collapsed emission is shown with contours $\pm(2,3,4,\dots)\sigma$. The synthesized beam is shown by the ellipse to the lower left. In the centre and right panels, the normalized radial profile of the intensity map is shown in brown with $1\sigma$ error bars, while that of the PSF is shown in cyan. The grey shaded region depicts a normalized mean intensity of $<0.5\times$(RMS noise level of the intensity map).}
\label{contstack_fig}
\end{figure*}

Since many of these sources are not detected in FIR continuum emission, we may only consider the `None' and `InvV' weighting schemes. Of these two, `InvV' shows a more significant divergence from a point-source profile. Again, we apply the fitting routine detailed in Section \ref{HALOMOD} to the stacked continuum map, resulting in the fits shown in Figure \ref{contstack_fig} and the parameters listed in Table \ref{pmntable}. We find that a point source returns the best fit, implying that the emission is unresolved. Using the average redshift of this sample ($z\sim2.3$), this implies an upper limit on the continuum size of $\lesssim0.6''\sim5$\,kpc.

A separate analysis of six FIR-detected SUPER sources \citep{lamp21} found a mean FIR size of $R_e=1.16\pm0.11$\,kpc. The observations in this previous work featured a higher spatial resolution ($\sim0.2''\sim2$\,kpc) and sampled a brighter region of the FIR SED (i.e., ALMA band 7 rather than band 3), so it is encouraging that our results are in agreement. This highlights the utility of high-resolution continuum observations for characterizing the morphology of the host galaxy. 

\subsubsection{Summary of stacking results}\label{stacksum}
Thus far, we have analysed the spatial extent of cold gas (traced by CO(3-2)), the stellar component (traced by HST/ACS i-band), and dust (traced by rest-frame FIR continuum emission). When the CO(3-2) emission is stacked in three-dimensional space and then collapsed in a small velocity bin (i.e., [-100,+100\,km\,s$^{-1}$]), we find strong evidence for two components: a compact ($r\sim1$\,kpc) centre and an extended ($r\sim13$\,kpc) but much weaker (i.e., peak amplitude $\sim0.005$ that of the centre) component. A similar fit is found if we first create CO(3-2) moment zero maps of each source and then stack them.

A different behaviour is found for the stellar distribution, as the radial profile of the high-resolution HST stack may be decomposed into a point source, a `bump' at $r\sim3$\,kpc, and a third component that extends to a maximum radius of $r\sim5-10$\,kpc. The complex stellar distribution may indicate small-scale merger or outflow activity that is not detectable in the relatively low-resolution ALMA observations. Alternatively, this could be a result of dust screening (e.g., \citealt{fuda21}). Since the CO emission is found to feature a HWHM$\sim13$\,kpc (not a maximum radius), the extended gaseous component has a much larger spatial extent than the stellar component. The difference in radial extents indicates that the extended CO emission is separate from the young stellar component, and thus may trace the inner CGM rather than the ISM. Alternatively, this extended reservoir could represent an extended disk or infalling gas from past outflows or a diffuse large-scale filament (see Section \ref{howform}).

Finally, the dust emission is best-modeled by a unresolved central source, with little evidence for an extended component. This stack includes 24 sources (both continuum-detected and continuum-undetected), making it sensitive to weak emission. Despite this, it only shows two low-significance extensions that are likely noise.

When combined, these analyses indicate a multi-layered structure of AGN host galaxies at Cosmic Noon: a compact core of dust, stars, and cold gas ($r\lesssim1$\,kpc), a complex stellar distribution that extends to $r\sim3-10$\,kpc, and an underlying diffuse gas reservoir with HWHM$\sim13$\,kpc. This agrees with previous observations of SUPER sources, which found half-light radii of $r\sim1-4$\,kpc for ionized gas (\oiii$\lambda$5007) and star formation (H$\alpha$; \citealt{lamp21}). The bulk motion of high-velocity ionized outflows in these sources were detected out to $r\sim3$\,kpc (\citealt{kakk20}). Other works found that high-mass star-forming galaxies around cosmic noon should feature stellar radii of $r_{80}\sim5$\,kpc \citep{mosl20}. Since the $r\sim13$\,kpc CO emission is beyond the stellar, dust, and star forming regions, we posit that it represents an extended gaseous reservoir.

%The current ALMA CO(3-2) data is insufficient to decipher the nature of the stellar distribution, so further high-resolution ($<1''$) ALMA observations are required. 

%We note that the values of log$_{10}$(HWHMW$_{G2}$/kpc) for the `cube stack', `moment 0 stack', and `continuum stack' agree to within $1\sigma$. However, the low Bayes factors of the latter two fits suggest that they are truly unresolved sources, and that this is a coincidence.

\subsection{How was the extended gaseous reservoir formed?}\label{howform}
The extended CO emission revealed by this stacking procedure may be caused by multiple possibilities: a number of discrete satellites, a low-level enriched gaseous reservoir, or single gaseous filaments (see discussion of \citealt{fuji19}). 

In order for the satellite explanation to be viable, the stellar contribution of these companions should contribute to the stack of stellar emission, and so we should also see weak extended emission in the F814W stack. We do not (Section \ref{hstsect}), so we conclude that satellites have little contribution to the diffuse emission.

Instead, it is possible that the extended CO emission represents a gaseous reservoir created by outflowing gas. Indeed, a detailed analysis of SUPER galaxies as observed in \oiii$\lambda$5007 line emission with SINFONI \citep{kakk20} found evidence for outflows of ionized gas with high velocities ($\sim650-2700$\,km\,s$^{-1}$). Assuming the redshift-dependent stellar mass-halo mass relation of \citet{gire20}, the SUPER galaxies should reside in halos of mass $\sim10^{12-13}$\,M$_{\odot}$. Thus some of this outflowing gas detected in SINFONI observations is travelling faster than the escape velocity ($\sim10^3$\,km\,s$^{-1}$ for a $\sim10^{13}$\,M$_{\odot}$ halo; \citealt{mill18}). Some of this gas may escape the halo, while the rest could be deposited into a gaseous reservoir or fall back into the galaxy (e.g., \citealt{spit13}).

While the outflows feature high enough velocities to escape the galaxy, it is not clear if they are sufficient to create an extended CO component with  HWHM$\sim13$\,kpc at $z\sim2.0-2.5$ (t$_{\mathrm{H}}=3.2-2.6$\,Gyr, respectively). As a sanity check, in the extreme case that the outflows began at Cosmic Dawn ($z\sim15$; e.g., \citealt{wise14}), the minimum outflow velocity required for the gas to reach this radius is only $<10$\,km\,s$^{-1}$, which is much less than the currently observed ionized outflow velocities. Instead, we may consider how long it would take the gas to reach the observed radius, considering a mean outflow velocity of 1000\,km\,s$^{-1}$. In this case, only $\sim13$\,Myr would be required. In reality, AGN outflows are expected to be highly episodic (e.g., \citealt{hick14,sun14}), so an assumption of constant velocity is not physical. But we are not able to rule out the possibility that outflows populated the extended component with molecular gas.

Cosmological simulations showed that bright galaxies at high redshift were likely encased in diffuse filaments of gas (e.g., \citealt{pall17,koha19}). In this case, the detected extended emission would not represent gas that emerged from the galaxy, but instead would signify infalling gas from the circumgalactic environment. However, this gas is expected to have densities $\lesssim3$ orders of magnitude lower than the central galaxy, making its detection difficult. It is also expected to have very low metallicity, and thus would be hardly visible in CO emission. 

Finally, it is possible that these galaxies are undergoing `inside-out growth' (e.g., \citealt{fran19}), where the star formation rate decreases with radius. The extended CO emission detected here would then represent a gaseous reservoir surrounding the galaxy that will contribute to future star formation. But since a CO detection implies that the gas is not pristine, the CO-traced gas must have had some past star formation, making this scenario less likely. This type of galactic feature is better detected in HI maps (e.g., \citealt{luce15}).

In summary, we find it unlikely that the extended emission is caused by a gaseous filament or individual satellite galaxies. The observed outflows feature sufficient velocities to populate an enriched gaseous reservoir at the observed radius. Additionally, the emission could be caused by an extended reservoir with a a low level of past star formation.

\subsection{Gas Mass}
Ideally, these results would be used to calculate the gas mass of the host galaxies and their associated extended components by converting L'$_{CO}$ to M$_{H_2}$ (see review by \citealt{bola13}). However, there are several issues that make this currently impossible for our dataset. 

Primarily, the conversion of CO(3-2) flux density to molecular gas density is dependent on two values: the ratio of L'$_{CO(3-2)}$/L'$_{CO(1-0)}\equiv r_{31}$ and the ratio of M$_{H_2}$/L'$_{CO(1-0)}\equiv\alpha_{CO}$. A common approach is to take an r$_{31}$ value for the galaxy type from literature (e.g., \citealt{cariw13}) and assume either a metallicity-based $\alpha_{CO}$ (e.g., \citealt{nara12}) or the common values for starbursts and Milky Way-like sources ($\sim0.8$ \& $\sim4.3\,$M$_{\odot}\,$K$^{-1}\,$km$^{-1}\,$s$\,$pc$^{-2}$ respectively; e.g., \citealt{bola13}). While values of $r_{31}$ for integrated CO luminosities only vary by $\sim0.5$\,dex, local observations have revealed that this quantity may vary by an additional $\sim0.25$\,dex between the central value and that at $r_{eff}$ (e.g., \citealt{lero22}). Since this value is truly dependent on the temperature and density of the gas (e.g., \citealt{vand07}), it is reasonable to expect that it should differ greatly for the host galaxy and the much less dense, colder CGM. Similarly, observations and simulations suggest that $\alpha_{CO}$ may vary by up to an order of magnitude within a galaxy (e.g., \citealt{teng22,hu22}), as it is dependent on metallicity, density, and optical depth (\citealt{madd20}). 

Because $r_{31}$ and $\alpha_{CO}$ have not been derived for the CGM of AGN host galaxies at cosmic noon, we are not able to place precise constraints on M$_{H_2,CGM}$. As a very basic estimate of the host galaxy gas mass, we may consider a representative L'$_{CO(3-2)}\sim10^{10}$\,K\,km\,s$^{-1}$\,pc$^{2}$ \citep{circ21}, $r_{31}\sim0.97$ \citep{cariw13}, and $\alpha_{CO}\sim0.8\,$M$_{\odot}\,$K$^{-1}\,$km$^{-1}\,$s$\,$pc$^{-2}$ (appropriate for high-\textit{z} QSO host galaxies; \citealt{deca22}). This results in log$_{10}(M_{H_2,Host}$/M$_{\odot})\sim9.89$. Taking the best-fit FWHMs and relative peak intensity from Table \ref{pmntable}, we find that the integrated flux density (S$\Delta v_{CO}$) of the extended component is $\sim2\times$ that of the host galaxy. Using the Milky Way-like values of $r_{31}=0.27$ (\citealt{cariw13}) and $\alpha_{CO}\sim4.3$ (\citealt{bola13}), we find log$_{10}(M_{H_2,CGM}$/M$_{\odot})\sim10.4$. However, we emphasize that this value is highly uncertain because of the issues discussed above.

\subsection{Comparison to other detections of extended gas emission}
Studies in recent years have revealed a number of high-redshift sources with extended \cii emission by using radial brightness profiles. Both \cii and CO are used as tracers of cold gas (e.g., \citealt{bola13,zane18}), so their spatial extension may be used as evidence for a gaseous reservoir. However, the interpretation of \cii is complicated by a strong dependence on gas conditions (e.g., \citealt{madd20}) and a variety of emission media (e.g., molecular and atomic gas; \citealt{pine13}). Here, we outline some of these discoveries and compare the results with our findings.

The first study to show evidence for high-\textit{z} extended gas components from ALMA \cii observations was \citet{fuji19}, who studied 18 galaxies at $z\sim5-7$. Each dataset was trimmed to only include [$-50,+50$]\,km\,s$^{-1}$ from $z_{[CII]}$, the visibilities were combined, the weighting was adjusted using the CASA task \textlcsc{statwt}, and a final stacked image was created. A radial profile from this image shows signal out to $\sim10$\,kpc and evidence for two components (i.e., host galaxy and extended gas reservoir). But since beam convolution is not taken into account when analyzing the radial profile, it is difficult to say what the intrinsic extension of the \cii emission is. While the small velocity range was selected to minimize the effects of close companions, the sample still includes two sources with close companions at the same velocity (HZ8 \citealt{capa15} which is also known as DEIMOS\_COSMOS\_873321 \citealt{beth20,jone21}, and WMH5 \citealt{jone17}), making the origin of the extended emission ambiguous (i.e., an extended component or close companions).

The study of \cii emitters at high-$z$ was greatly progressed by the ALPINE survey (\citealt{lefe20,beth20,fais20}), which used ALMA to observe \cii emission from 118 SFGs between $z\sim4-6$ at $\sim1''$ resolution ($\sim6-7$\,kpc). \citet{fuji20} examined the \cii moment maps and radial brightness profiles of the 46 individual sources that were detected at $S/N_{[CII]}>5$, claiming evidence for extended emission to $\sim10$\,kpc. However, as before, if the beam convolution is taken into account, the true sizes are only $r\sim1-3$\,kpc. Because of this, it is unlikely that the individual sources exhibit reliable evidence for true extended gas components, based only on ALPINE data.

In a separate analysis, \citet{gino20} examined the 50 ALPINE galaxies that were detected at $3.5\sigma$ and showed no signs of merging (according to the morpho-kinematic classification of \citealt{lefe20}, but see further discussions of morpho-kinematic classification difficulty for low-resolution data in ALPINE: \citealt{jone21,roma21}). This sample is further split into galaxies with SFR$<25$\,M$_{\odot}$\,year$^{-1}$ (30 objects) and SFR$>25$\,M$_{\odot}$\,year$^{-1}$ (20 objects; based on ALPINE DR1\footnote{\url{https://cesam.lam.fr/a2c2s/dr1}}). The data cubes of the high-SFR subsample are stacked in three dimensions in a nearly identical way to this work, and a moment 0 map is made for [-200,+200]\,km\,s$^{-1}$. A radial profile of this map reveals a distinct discontinuity, with a compact inner component and a low-level extended component with an intrinsic radius of $\sim10$\,kpc. This extended component is likely created by SF-driven outflows, as the low-SFR subsample does not show evidence for such a component and the ALPINE sample was chosen to exclude type 1 AGN.  

The ALPINE survey has now been followed by the REBELS survey, which is observing \cii from 40 SFGs at $z\sim6.5-9.0$ \citep{bouw21}. The spatial extent of \cii emission in 28 well-detected galaxies in this survey has been examined by \citet{fuda22}. By stacking moment 0 maps and extracting radial profiles, they find little evidence for two spatial components, as the emission is well fit with a single component with $r_e\sim2.2$\,kpc. This analysis is repeated with ALPINE data, revealing a similar result for the $z\sim4-6$ galaxies. The lack of an extended component may be due to the stacking of moment maps rather than data cubes (see discussion in Section \ref{stacksum}), the inclusion of both high- and low-SFR sources, or may be a intrinsic lack of extended emission. However, in all cases the \cii emission was found to be $\sim2\times$ as extended as the rest-frame UV and FIR dust continuum emission. 

The $z\sim5.5$ SFG HZ4 was observed at high resolution ($0.3''\sim2$\,kpc) and high sensitivity (0.15\,mJy\,beam$^{-1}$ RMS noise level in 16\,km\,s$^{-1}$ channels) in \cii emission by \citet{herr21}. This source is similar to those of the ALPINE sample (i.e., a $z\sim4-6$ SFG with no evidence for type 1 AGN), and the high-resolution observations yielded a similar extended \cii brightness profile with an intrinsic \cii radius of $r\sim6$\,kpc , which extended past the dust and UV emission. 

At slightly higher redshift, \citet{meye22} observed the $z=6.42$ QSO J1148+5251 in \cii emission with ALMA. The resulting radial brightness profile revealed extended emission with an intrinsic radius of $\sim10$\,kpc.

Very recently, \citealt{akin22} analyzed the spatial distributions of \cii, \oiii88$\mu$m, FIR dust, and rest-frame UV emission in the $z=7.13$ SFG A1689-zD1. While the \oiii88$\mu$m and UV profiles show single, resolved components, the FIR and \cii both show evidence for two components. This extended emission is significant out to a convolved radius of $\sim12$\,kpc, but the deconvolved HWHM of the extended emission is not apparent. In addition, the source features two spatial peaks in UV continuum emission and a disturbed morphology in the other tracers, suggesting either an ongoing merger or a `clumpy' morphology. The interpretation of this extended \cii emission is thus made more difficult.

At lower redshifts, CO transitions have also be used to investigate extended emission around a few high-z galaxies. \cite{Ginolfi17} mapped CO(3-4) in a galaxy at z=3.5 and found evidence for emission extending to a radius of about 20 kpc, well beyond the stellar disc. CO observations of powerful radio loud AGNs have found extended molecular gas on even larger scales of several tens of kpc \citep{Emonts16,li21}.

ACA observations of one of the SUPER galaxies in CO(3-2) emission revealed evidence for a large amount of molecular gas out to $r\sim200$\,kpc (CID\_346; \citealt{cico21}), implying an enormous history of outflows that enriched the CGM out to large radii. Unfortunately a deeper ACA CO(3-2) observation of this source does not reveal evidence for any significant emission beyond $r\sim10$\,kpc (Jones et al. submitted). However, the concept of outflows creating such a large gaseous reservoir is intriguing. As noted by the authors, a definitive detection of this gas component will require a new instrument (e.g., AtLAST; \citealt{klaa20}).

To summarize this past work, a number of \cii and CO observations of higher-redshift ($z\sim3-6$) SFGs and a QSO show that the individual sources exhibit larger extensions in \cii than rest-frame UV (\citealt{Ginolfi17,fuji20,herr21,meye22}). On the other hand, a stack of high-SFR SFGs (\citealt{gino20}) and a strongly lensed $z\sim7$ SFG show evidence for a second spatial component on a scale of $\sim10\,$kpc. This is the same scale as we find in this work for a stack of CO(3-2) AGN host galaxies at cosmic noon ($z\sim2$). 

While the comparison sample features galaxies with different outflow-driving engines (i.e., star formation winds rather than AGN feedback) and is in a different cosmological era (i.e., $t_H\lesssim1.5$\,Gyr rather than $t_H\sim3$\,Gyr), the common size scale of the extended components suggests a link of some sort. Indeed, the ALPINE sources and a number of SUPER host galaxies are located on the main sequence for their redshift (i.e., they are neither starbursts or quiescent galaxies), which may indicate a similar evolution. Since enriched reservoirs could either be created by AGN or starburst-driven outflows (e.g., \citealt{maio12,fior17,spil18,Fluetsch2019,jone19,schn20,Lutz20}), both AGN and starbursts are episodic events (e.g., \citealt{nova11,arat20,talb22}), and some galaxies host both starbursts and AGN (e.g., \citealt{sand96,bane17}), these reservoirs were likely created by a combination of past AGN and starburst activity for all sources.

\subsection{Stacking Assumptions}\label{stas}
Throughout this work, we have utilized image stacking to increase the S/N of our data and explore low-brightness emission. Stacking is a powerful tool that allows us to increase the S/N of our data without requiring more observing time. But its use rests on several assumptions. 

\textbf{Sample Uniformity:} Primarily, the input sample is assumed to be relatively uniform in intrinsic properties (e.g., size, luminosity), observational characteristics (e.g., beam size, RMS noise level), and imaging parameters (e.g., cell size, phase centre). 

By adopting a combined stack weighting scheme of inverse variance, we are able to account for variations in intrinsic RMS noise level, but not CO line strength. Since our goal is to search for extended emission, it is reasonable to allow strong sources to dominate the stack. Our imaging pipeline ensures that the cell size of each image is uniform, while the beam size and phase center of each data cube is similar since the stacked data was taken from the same observing programs. While the beam properties (major axis length, axis ratio, and position angle) do vary slightly between data cubes (see Table \ref{basictab}), we account for this by stacking the beams of each cube in the same way as the data.

\textbf{Physical Scale:} Our CO(3-2) sample includes sources in the redshift range $z=2.219-2.450$, implying a angular scale of $8.253-8.105''$/kpc. This $\sim2\%$ spread has no serious effect on our conclusions. Similarly, while we find that the HST/ACS i-band spatial centroids of each source may differ from the ALMA CO(3-2) positions (\citealt{circ18,circ21}) by up to $\sim0.24''$ (or $\sim1$ pixel in the CO cubes), this has little effect on the large-scale (i.e., $>1''$) emission that we detect.

\textbf{PSF Sidelobes:} The above approach does not alleviate all of the issues associated with analysing interferometer-based data in the image plane. The true PSF of these data is not just a 2-D Gaussian, but features complex variations over  the field of view (i.e., sidelobes of $\lesssim10\%$). The cleaning process (i.e., CASA \textlcsc{tclean}) acts to reduce the effects of these sidelobes by iteratively identifying the brightest pixel, convolving it with the true PSF, and subtracting this from the original (`dirty') map. However, this does not remove the sidelobes entirely, as it only cleans down to a user-provided cleaning threshold, which is usually set to $3\times$ the RMS noise level (e.g., \citealt{spil16,bisc19,gino20,yu21})\footnote{Although other studies use a threshold of $1.5-2.0\sigma$ (e.g., \citealt{kaas20,alge21,garc22}).}. While a lower threshold (e.g., $0\sigma$) would ensure that the effects of the sidelobes are minimized, this would introduce a number of artifacts due to noise peaks. On the other hand, a higher threshold (e.g., $5\sigma$) would ensure that there are no artifacts caused by noise, but also would result in high-level sidelobes. Since we use a threshold of $3\sigma$ (see Section \ref{obsdat}), it is possible that we have not fully subtracted the sidelobes of the PSF.

This is of importance for our work because we wish to search for low-level ($\lesssim10\%$ of peak) at large spatial separations from our sources. Luckily, there are two properties of our analysis that minimize the effects of these sidelobes. First, an inspection of the PSFs show that they are not azimuthally symmetric Airy disks, but feature some isolated peaks and troughs. Second, our stacking analysis smears these PSFs together, minimizing the effects of local extrema. By then extracting radial profiles, the effects of the sidelobes are again lessened. 

While these factors result in a greatly lessened effect of sidelobes in our analysis, we note that the best way to ensure their absence is to switch from the image plane to visibility-based analysis (e.g., \citealt{fuji19}, Scholtz et al. submitted). 

\textbf{Visibility or Image Stacking}: Working in the visibility plane has several strengths over the image plane, as there is no need to consider weighting schemes (e.g., natural, uniform, Briggs `robust'; \citealt{brig95}), cell size, or cleaning parameters (e.g., threshold, iterations, primary beam limit). In short, the visibilities are the ``pure'' data seen by the interferometer without any further (Fourier) transformation, processing or interpolation of the data, and are hence more reliable for detecting and characterising (simple) signals (e.g., \citealt{fuji19,fuda22}).

But only simple morphologies may be assumed (generally only point sources or Gaussians) and faint companions (which appear as low-level sinusoidal contributions to the real part of the visibilties) may be difficult to detect. This is where image-plane stacking is useful, as we may examine the full 2-D distribution of flux and search for discrete emission that may signal a satellite galaxy.

With this in mind, image- and visibility-plan analyses are complementary. While we will utilize the latter method in future works to analyze the distribution of emission (Jones et al. submitted, Scholtz et al. submitted), we posit that the beam-deconvolved analysis in this work is self-sufficient.

\section{Conclusions}\label{conc}
In this work, we have analyzed ALMA CO(3-2) data from a set of $z\sim 2.0-2.5$ AGN host galaxies from the SUPER survey, with the intent of searching for extended gaseous components that may hint at past outflows.
\begin{itemize}
    \item By performing a three-dimensional stacking analysis and extracting radial profiles of surface brightness, we find qualitative evidence for extended emission that deviates from the PSF.
    \item To examine these profiles in more depth, we create a brightness model that consists of one or two circular Gaussians, which we convolve with the observed PSF. These model profiles are then compared with the observed brightness profiles, and the intrinsic properties are fit for using the Bayesian inference code MultiNest. The single-Gaussian model returns an intrinsic (i.e., deconvolved) source radius of HWHM$=0.47^{+0.07}_{-0.06}"=3.89^{+0.58}_{-0.50}$\,kpc, while the two-Gaussian model returns a bright central source (HWHM$=0.09^{+0.15}_{-0.06}"=0.76^{+1.19}_{-0.46}$\,kpc) and a weaker extended component (HWHM$=1.66^{+0.58}_{-0.43}"=13.49^{+4.71}_{-3.49}$\,kpc). The two-Gaussian model returns a better goodness of fit (i.e., $\chi^2$, $\chi_{red}^2$, and Bayes factor). 
    \item A similar stacking and radial profile analysis is run on HST/ACS i-band SUPER galaxies, finding evidence for complex behaviour for $r\lesssim10$\,kpc but no emission at the CO-derived radius of $r\sim13$\,kpc. If the sample is split into NL and BL AGN, then this behaviour is present for both subsamples.
    \item We repeat the stacking analysis in two dimensions by using the CO moment 0 maps and rest-frame FIR continuum maps for the sample. For the former, we find similar results to the `cube stack' analysis, although at lower significance. The continuum emission is best fit by a point source, implying a size limit of $r\lesssim5$\,kpc.
    \item Various causes of this extended emission are explored, including enriched outflows caused by AGN feedback, satellite galaxies, and a background filament of low-density gas. Based on our analyses of the ALMA CO(3-2) and HST/ACS i-band emission, we conclude that the most likely case is that this emission represents processed material ejected from the host galaxy by past outflows. However, we note that other explanations (e.g., inside-out star formation) are also possible, and further data (e.g., high-sensitivity ACA observations) are required for a robust conclusion.
    \item Since it is not currently possible to convert the observed CO(3-2) luminosity in the extended component to a molecular gas mass, we discuss the assumptions that would have to be made.
    \item A comparison of our results to those of other high-redshift sources reveals a similar extent of molecular gas around a diverse sample of galaxies ($r\sim10$\,kpc). Due to the diverse sample properties (i.e., SFGs, AGN host galaxies, QSO host galaxies) and large redshift range ($z\sim2-7$), this is interpreted as evidence for a common series of star formation- and AGN-driven outflows over a large timescale.
\end{itemize}
Altogether, these analyses suggest that the molecular gas in SUPER galaxies exhibit a central peak, surrounded by an extended gaseous reservoir. That this extended component is not detected in a stack of the HST/ACS i-band images implies that it is caused by expulsion of gas by AGN feedback.

\section*{Acknowledgements}
We thank the referee for constructive feedback that has strengthened this work. This paper is based on data obtained with the ALMA Observatory, under programs 2016.1.00798.S, 2017.1.00893.S, and 2021.1.00327.S. ALMA is a partnership of ESO (representing its member states), NSF (USA) and NINS (Japan), together with NRC (Canada), MOST and ASIAA (Taiwan), and KASI (Republic of Korea), in cooperation with the Republic of Chile. The Joint ALMA Observatory is operated by ESO, AUI/NRAO and NAOJ. This research has made use of the NASA/IPAC Infrared Science Archive, which is funded by the National Aeronautics and Space Administration and operated by the California Institute of Technology. G.C.J. acknowledges funding from ERC Advanced Grant 789056 ``FirstGalaxies’’. R.M. and J.S. acknowledge funding from ERC Advanced Grant 695671 ``QUENCH’’ under the European Union’s Horizon 2020 research and innovation programme, as well as support by the Science and Technology Facilities Council (STFC). S.C is supported by European Union’s HE ERC Starting Grant No. 101040227 - WINGS. Y.F. acknowledge support from NAOJ ALMA Scientific Research Grant number 2020-16B.
 
\section*{Data Availability}
The data analysed in this work are available from the ALMA data archive (\url{https://almascience.nrao.edu/asax/}) and the IPAC COSMOS server (\url{https://irsa.ipac.caltech.edu/data/COSMOS/index_cutouts.html}).

\bibliographystyle{mnras}
\bibliography{references}

\appendix

\section{X\_N\_6\_27 Re-Analysis}\label{627sec}
Following the success of the SUPER-ALMA programs (2016.1.00798.S, 2017.1.00893.S; PI V. Mainieri), a new ALMA program was proposed to perform deeper observations of the CO(3-2) emission of the nine sources previously detected in \citet{circ21} (2021.1.00327.S; PI: R. Maiolino). These observations are ongoing, and will be presented in later works (Jones et al. submitted, Maiolino et al. in prep). 

However, new ALMA observations of X\_N\_6\_27 are complete, and feature a slightly larger synthesized beam size ($\sim2''$) compared to previous observations ($\sim1''$). The total on-source time is much longer ($\sim2.8$\,hours vs. $\sim9$\,minutes), resulting in a lower noise level.

The data from these observations were downloaded from the ALMA archive, and were calibrated using the scripts provided by ALMA staff. As a first step, we produce a data cube with no primary beam correction or cleaning. Emission is not detected at the original redshift ($z=2.2640$, corresponding to $\nu_{obs}=105.942$\,GHz). However, we do detect emission at $105.405$\,GHz ($z=2.2806$), or $\sim1500$\,km\,s$^{-1}$ from the redshift expected from previous spectroscopic redshifts. This offset is large, but comparable to the offset of other SUPER sources (e.g., $\sim1100$\,km\,s$^{-1}$ for CID\_166, $\sim1400$\,km\,s$^{-1}$ for X\_N\_44\_64).

\begin{figure}
    \centering
    \includegraphics[width=0.5\textwidth]{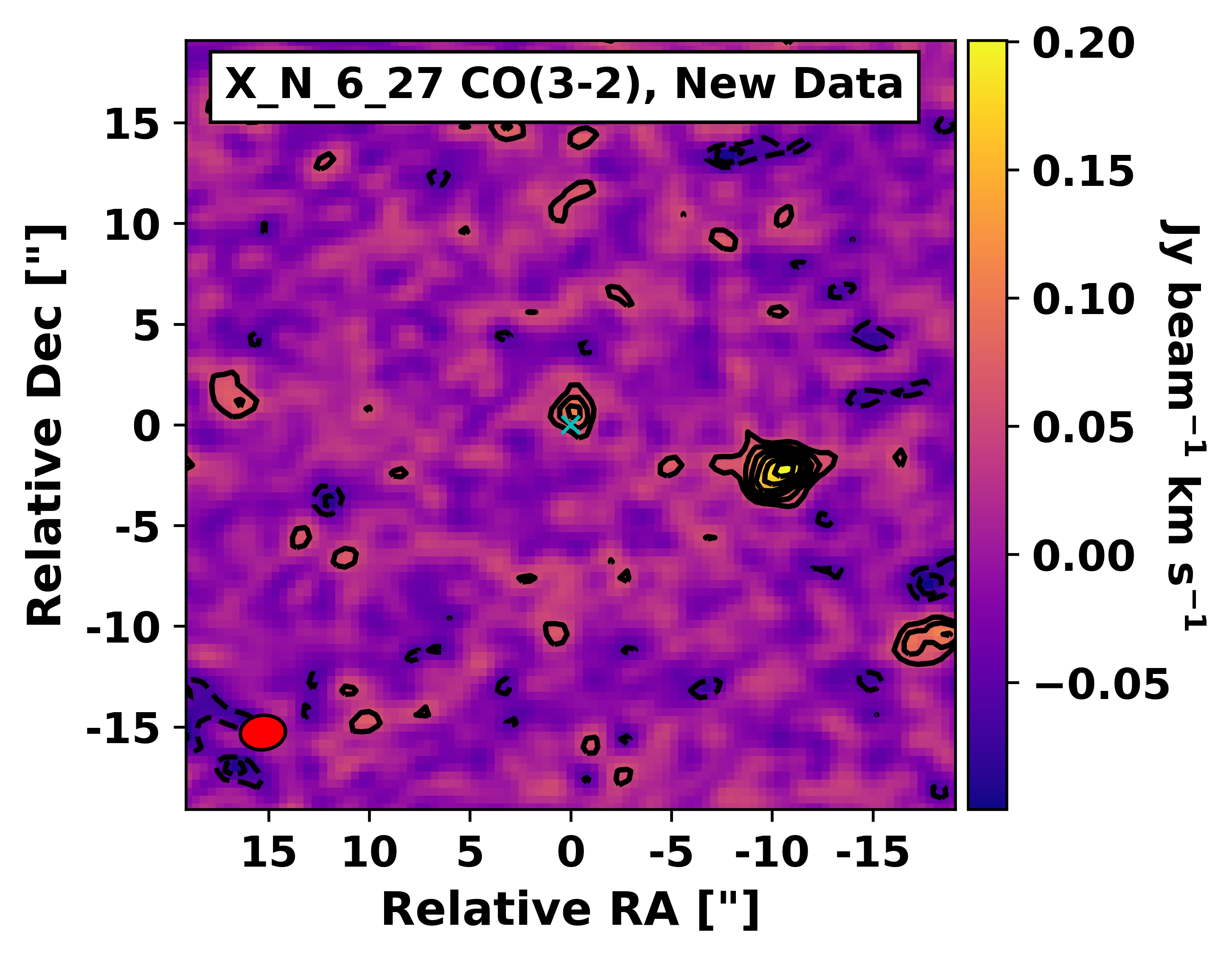}
    \includegraphics[width=0.5\textwidth]{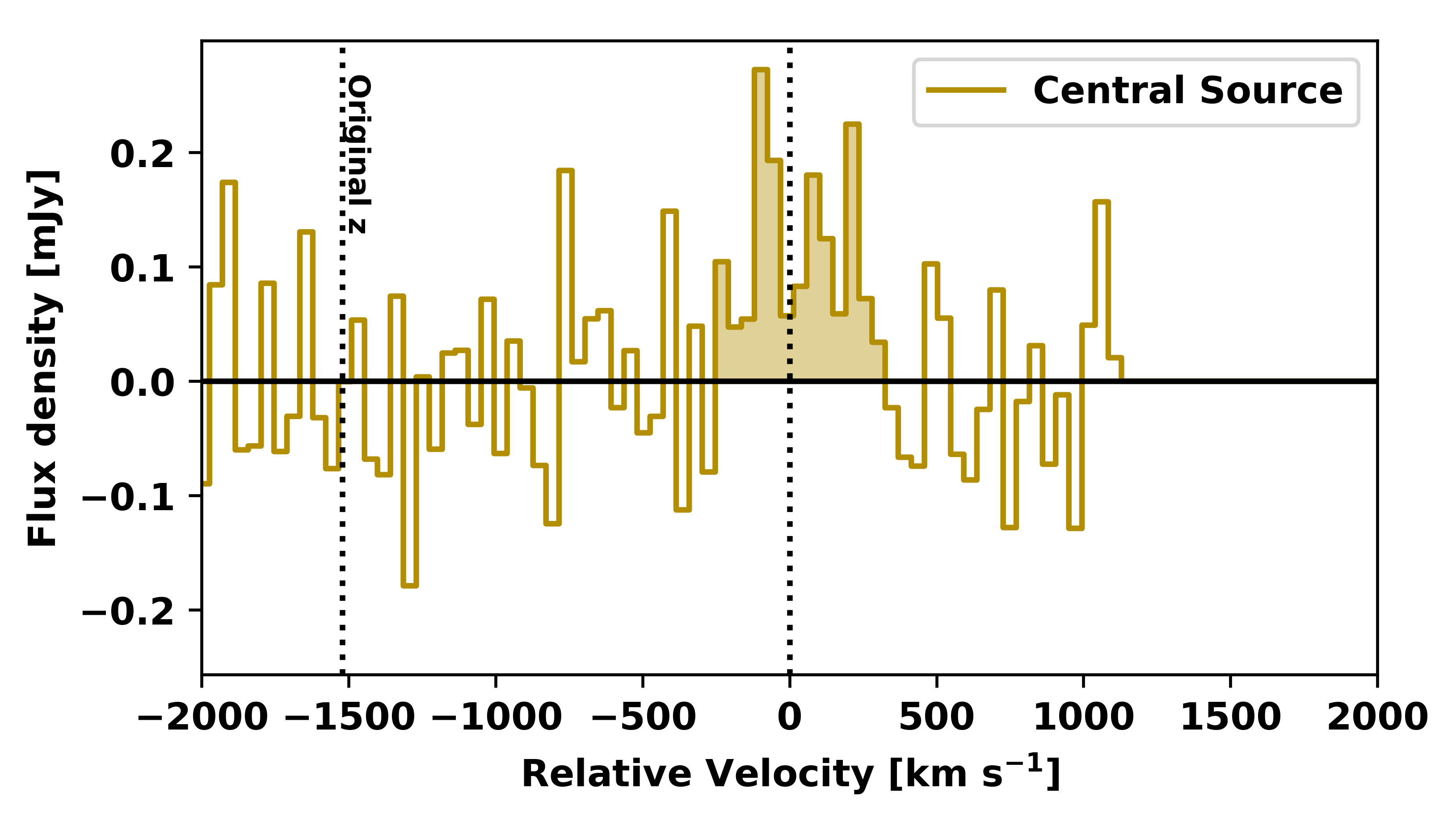}
    \includegraphics[width=0.5\textwidth]{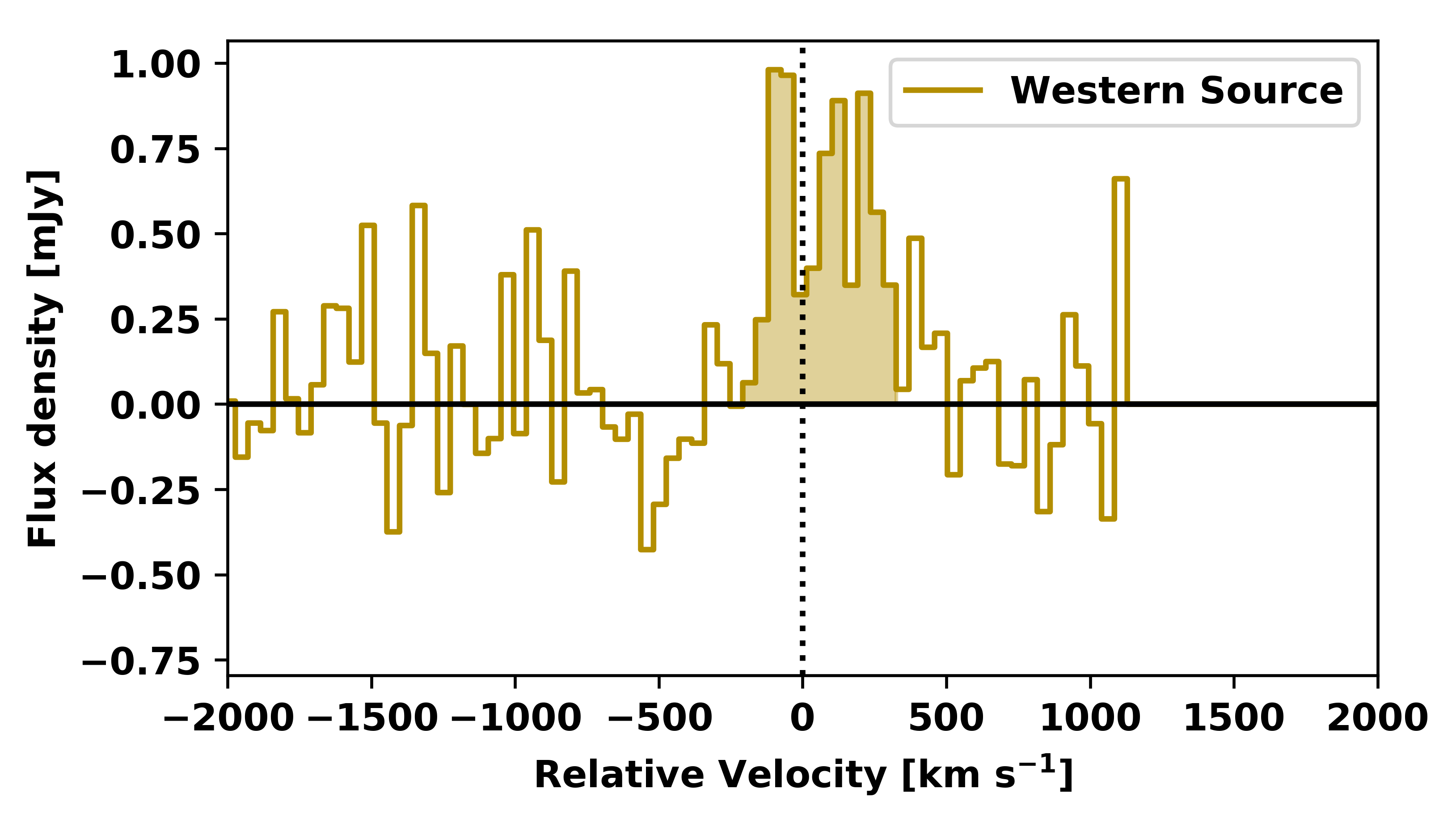}
    \caption{Initial results from further observations of CO(3-2) emission in X\_N\_6\_27 (Project 2021.1.00327.S). Emission is detected both at the phase centre and from a source to the west. The moment zero map of this emission is shown in the top panel, while the spectra of each source (using the $2\sigma$ contours as apertures) are shown below. Contours begin at $\pm2\sigma$ and are in steps of $1\sigma=0.027$\,Jy\,beam$^{-1}$\,km\,s$^{-1}$. The new redshift is shown as a vertical line at $v=0$\,km\,s$^{-1}$, while the originally reported redshift \citep{circ21} is shown at $\sim-1500$\,km\,s$^{-1}$. For both spectra, the collapsed channels are highlighted.}
    \label{newxn627}
\end{figure}

The resulting moment zero map and spectrum are shown in Figure \ref{newxn627}. A $\sim4\sigma$ source is present at the phase centre, while a much stronger source lies to the west. Both feature similar central frequencies and linewidths. The amplitude of the central detection is quite low ($\sim0.1$\,mJy), and would fall within the noise of previous observations ($\sim0.5$\,mJy, see figure D.1 of \citealt{circ21}). While further analysis of these sources (including a more thorough cleaning process) is deferred to a future work, we may state that the CO(3-2) emission line of this source is not detected in the original SUPER-ALMA data, so we do not include it in our analysis.

%---

\section{Uncertainty in radial surface brightness profiles}\label{uncanpp}
While it is now common practice to use radial profiles to characterize surface brightness distributions, there are slightly different approaches to determine the appropriate uncertainties in each radial bin. For example, some use the RMS noise level of the image (e.g., \citealt{meye22}), while others use the standard deviation of values in each annulus (e.g., \citealt{gino20,fuda22}). Others use bootstrap analyses to determine the background noise level (e.g., \citealt{fuji20}) or the variance of their sample (e.g., \citealt{fuji19}). It is also possible to present the Poisson noise (i.e., the RMS noise level of the image divided by the square root of the number of pixels in each annulus; \citealt{gino20}).

This variety of approaches arises from complexities of noise in interferometric images. Primarily, the noise per pixel is not independent, but is spatially correlated on the scale of the beam. Since annuli sample pixels from multiple beams, the appropriate uncertainity for the average value is not quite clear.

To determine the true behaviour of noise, we created a test image of 100$\times$100 pixels ($0.2''/$px) of pure Gaussian noise (noise level RMS$_1$), and convolved this with the 2-D Gaussian beam of the $\pm100$\,km\,s$^{-1}$, InvV-weighted moment zero map of Figure \ref{bigbig} ($1.38''\times1.24''$ at $0.82^{\circ}$). The convolved image exhibits a lower RMS noise level (RMS$_2$), so we adjust RMS$_1$ so that RMS$_2$ agrees with an observed value (RMS$_{\mathrm{M0}}$) to within $5\%$. Radial profiles of this convolved image are then taken, and the RMS in each annulus is calculated. This process is repeated 100 times.

\begin{figure}
    \centering
    \includegraphics[width=0.5\textwidth]{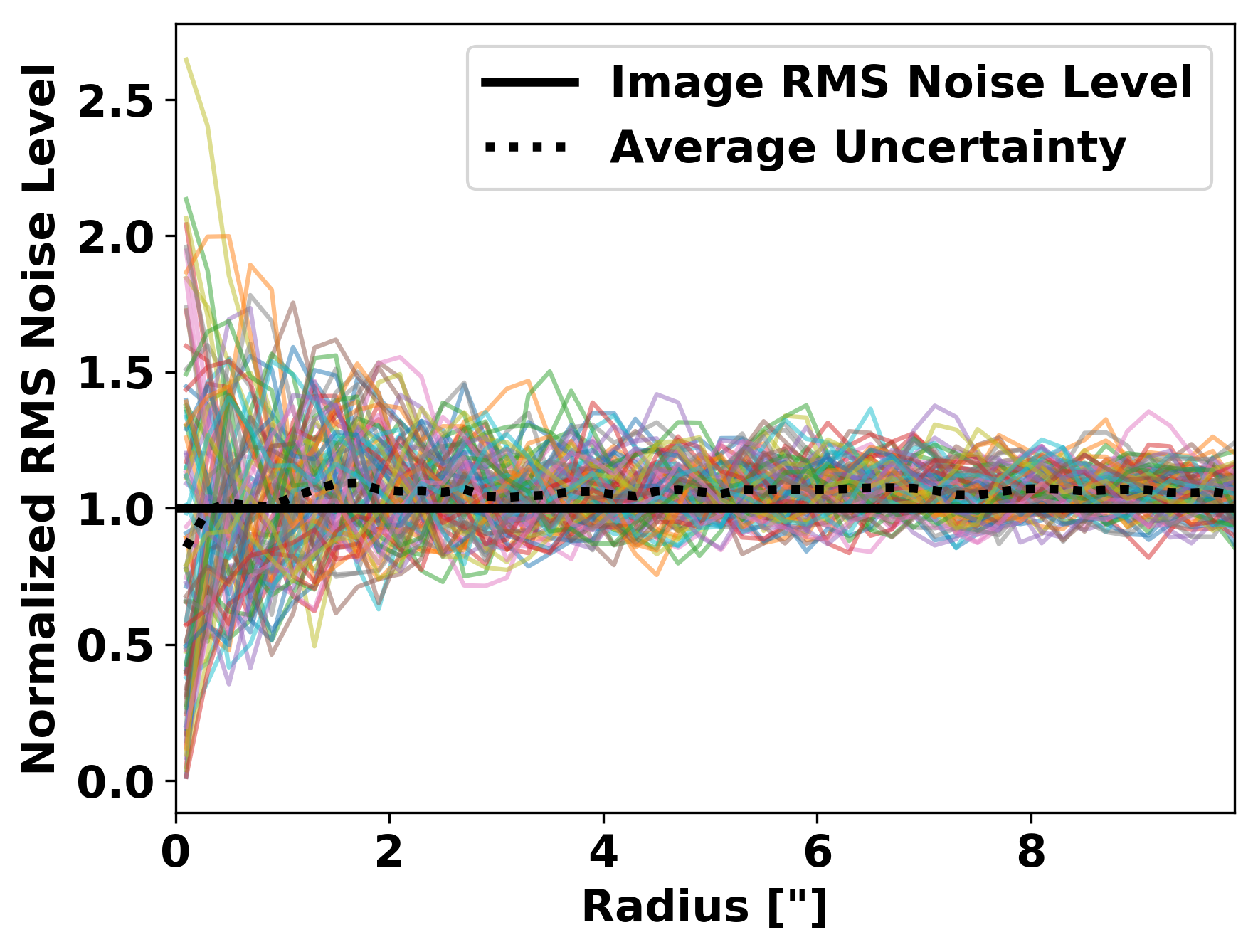}
    \caption{Results of radial profile uncertainty test. Colored lines show RMS noise level in annuli for 100 individual noise simulations. The average noise level in each bin is shown by the dashed black line. All noise levels are normalized to the RMS noise level of the image (solid black line).}
    \label{rmsimage}
\end{figure}

The resulting RMS profiles are depicted as colored lines in Figure \ref{rmsimage}. It is clear that the scatter in RMS decreases with increasing radius, due to the increasing number of pixels in larger annuli. However, the average RMS profile (dashed line) increases slightly at small radii before levelling off at a slightly higher value than RMS$_{\mathrm{M0}}$ (solid line). 

From this, it is clear that a Poisson error or standard error (which are dependent on the number of pixels in each annulus) are not appropriate. The RMS noise level of the image (RMS$_{\mathrm{M0}}$) is a reasonable approximation, as it lies within the scatter of RMS profiles. But our full treatment of the beam-dependent RMS per annulus is more precise.

We note that this analysis assumes a signal-free environment. In reality, the map will also include a source that is convolved with the beam, without azimuthal symmetry. To take this into account, the radial profile uncertainty in this work is the maximum of the beam-dependent RMS level (as above) and the standard deviation in the annulus. 

%---

\section{Covariance Plots}\label{corner}

\begin{figure}
    \centering
    \includegraphics[width=0.5\textwidth]{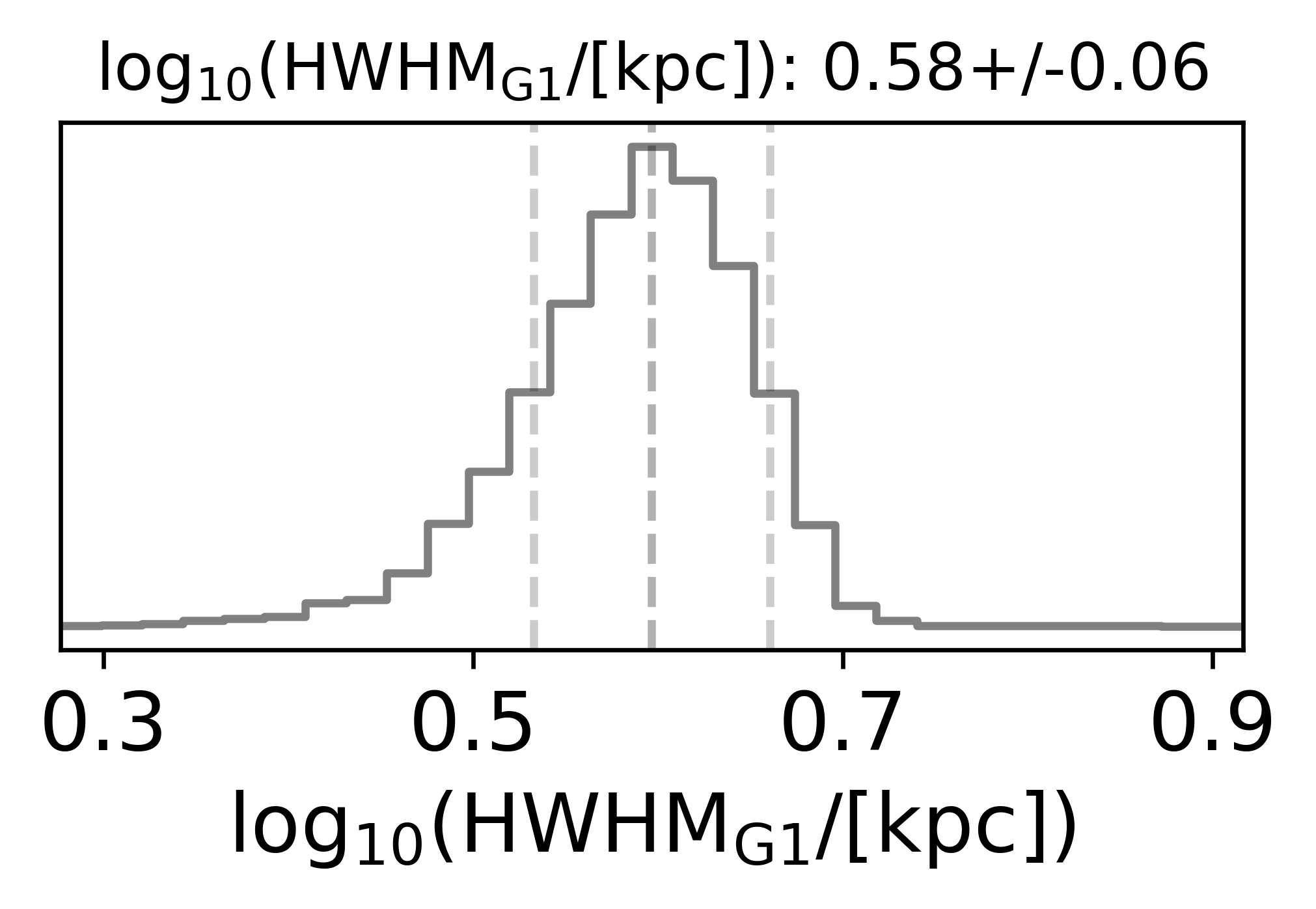}
    \caption{Posterior probability distribution for the best-fit intrinsic HWHM of the analyzed moment zero map when assuming a single 2-D Gaussian component (see top panel of Figure \ref{PMN_C}; Section \ref{HALOMOD}). Vertical lines correspond to the best-fit value and $\pm1\sigma$.}
    \label{post_1}
\end{figure}

\begin{figure*}
    \centering
    \includegraphics[width=\textwidth]{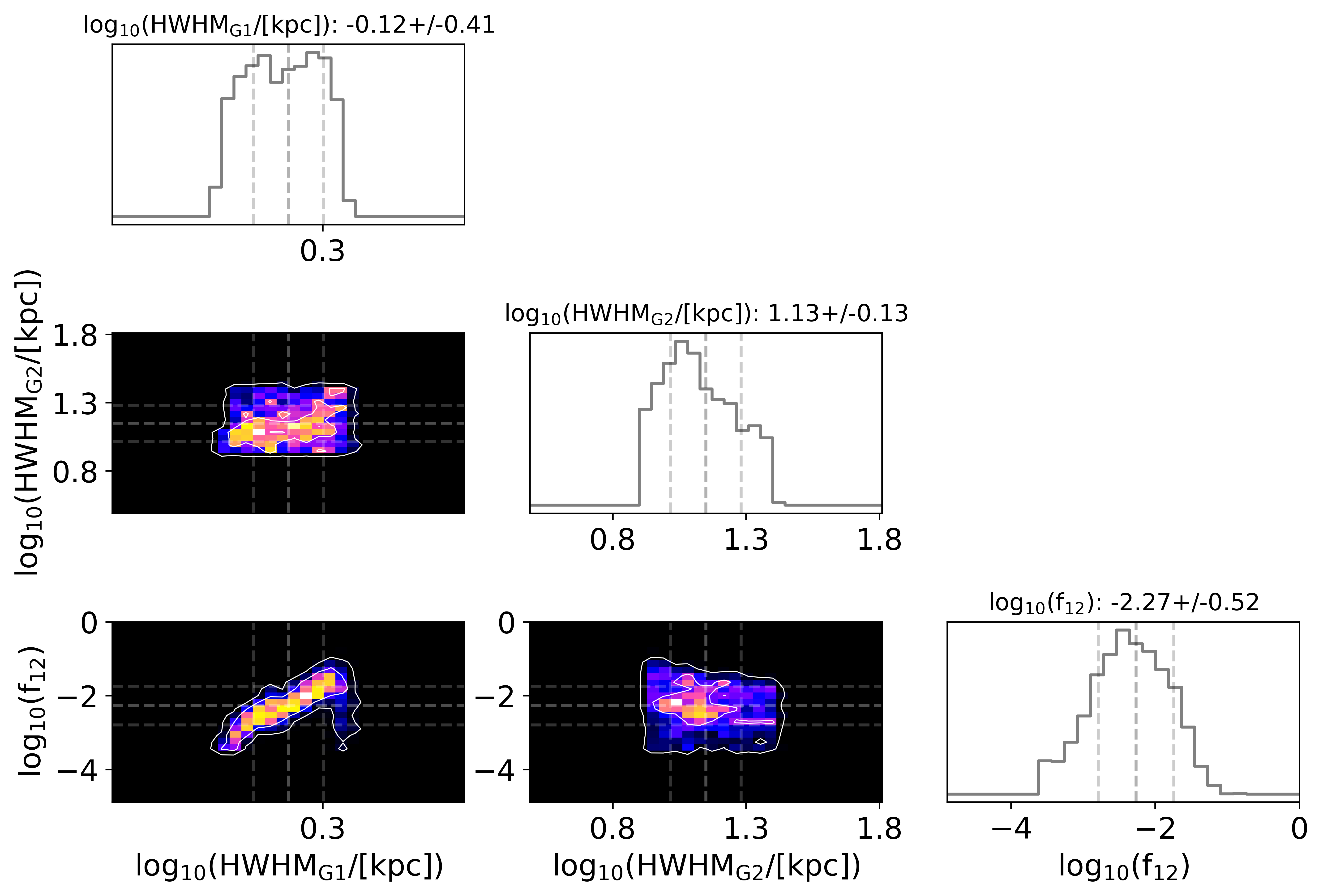}
    \caption{Corner plots for the best-fit source parameters when assuming two 2-D Gaussian components (see bottom panel of Figure \ref{PMN_C}; Section \ref{HALOMOD}). Top panels of each column show posterior probability distributions for each variable, while the coloured plots are covariance distributions. Vertical and horizontal lines correspond to the best-fit value and $\pm1\sigma$. }
    \label{post_2}
\end{figure*}

\section{Convolved HST Profile Test}
\begin{figure*}
\centering
\includegraphics[width=0.5\textwidth]{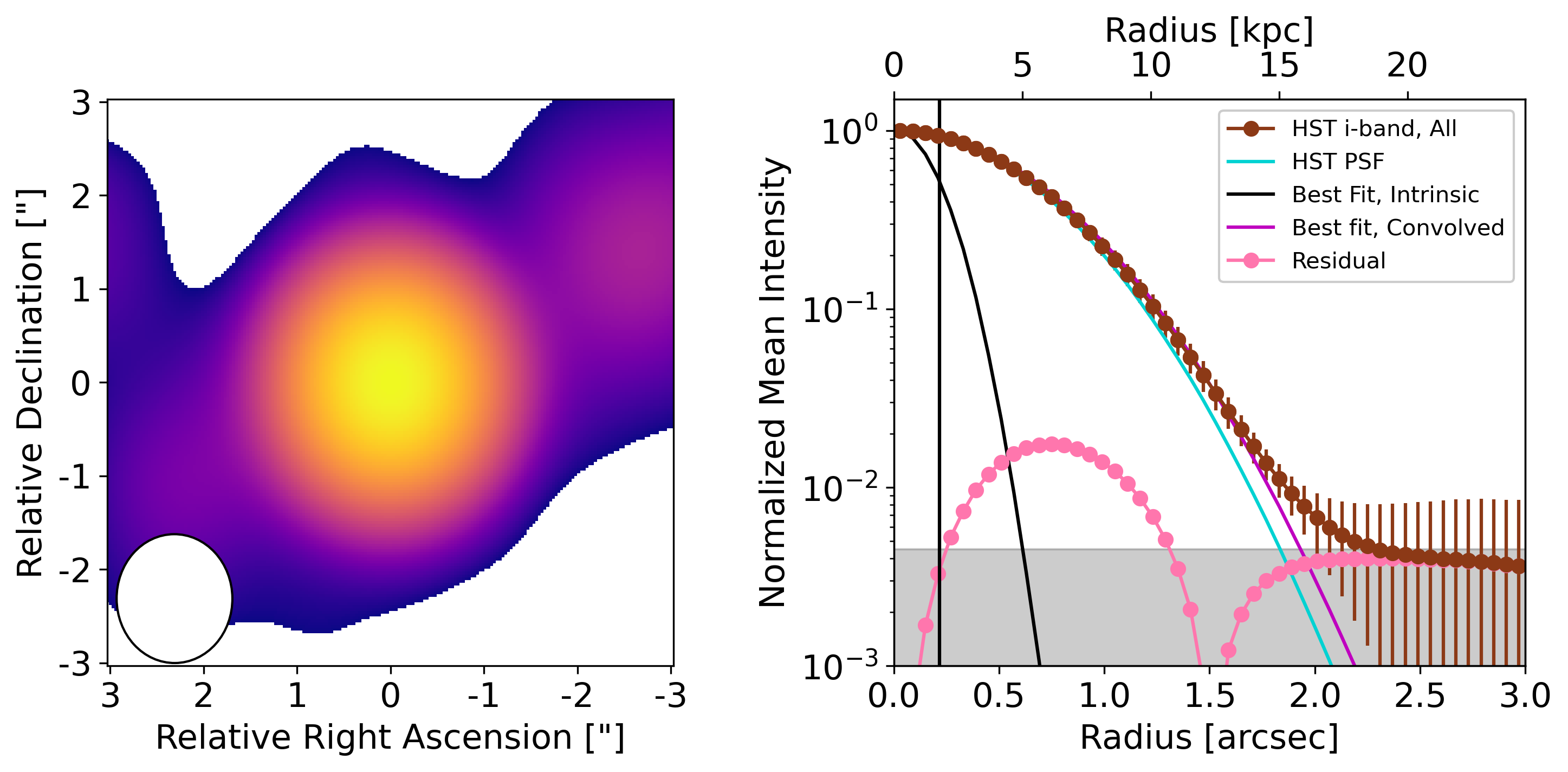}
\includegraphics[trim=11.6cm 0 0 0, clip, width=0.263\textwidth]{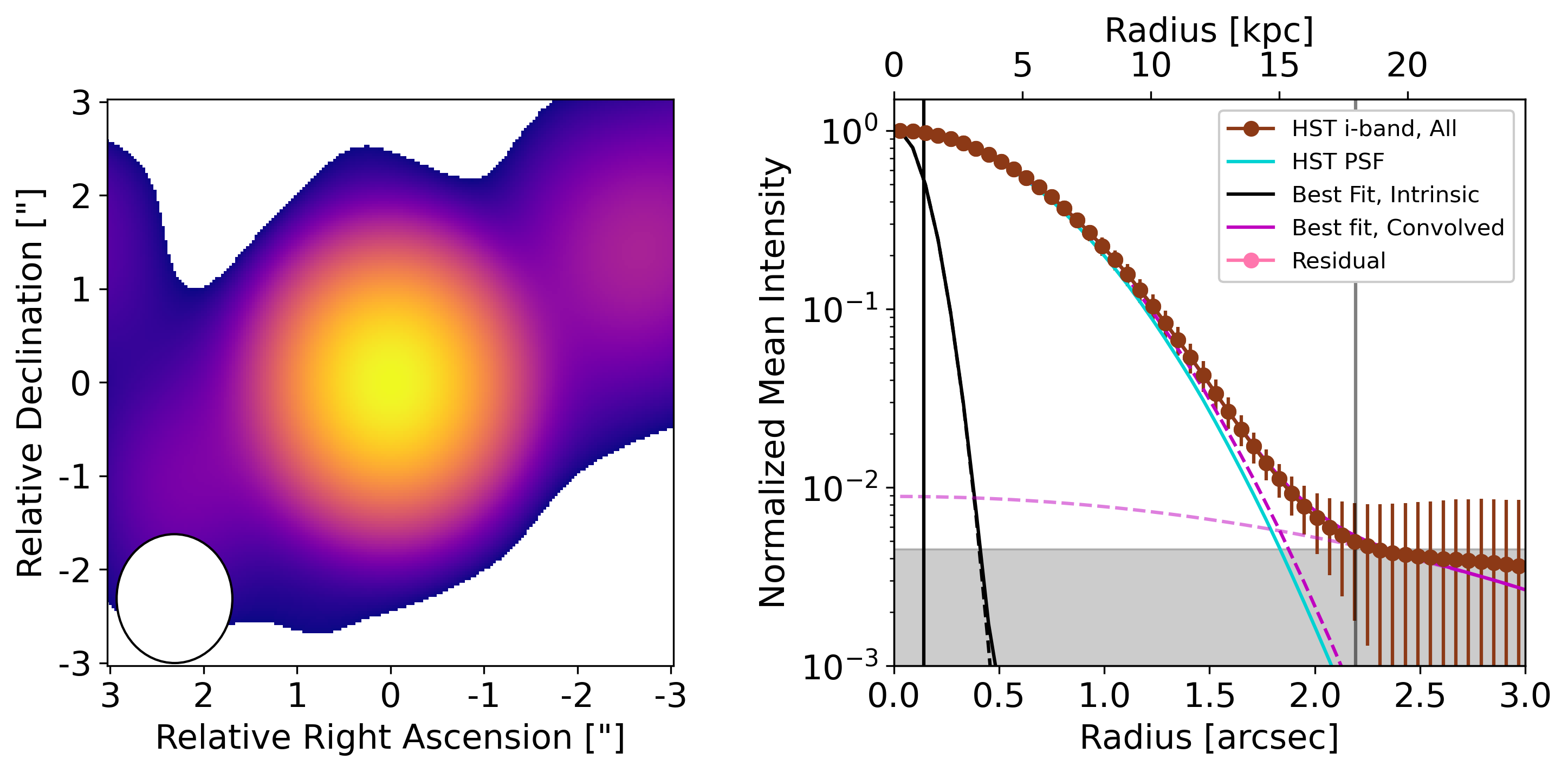}
\caption{Results of fitting models to the stacked image of HST/ACS i-band emission using 21 SUPER galaxies (see Section \ref{hstsect}). In each panel, the brown and cyan lines show the normalized mean radial profiles of the stacked map and beam, while the shaded region depicts $<0.5\times$(RMS noise level of the moment 0 map). For the one-Gaussian model (central panel), we present the best-fit model radial profile (magenta) and intrinsic (i.e., unconvolved) profile (black solid curves), as well as the intrinsic HWHM value of the best-fit model (vertical black line) . For the two-Gaussian model (right panel), the intrinsic and convolved components of the best-fit model are shown by dashed black and magneta lines, respectively. The vertical lines show the best-fit intrinsic HWHMs of each component.Note that in this case, the residual and intrinsic extended component lie below the lower y-limit. Each profile is normalized to its maximum value.}
\label{PMN_HST}
\end{figure*}

\label{lastpage}
\end{document}